\documentclass{JFM-FLM_Au}

\lefttitle{Xianglong Li et al.}
\righttitle{Journal of Fluid Mechanics}

\title{A Hybrid POD-Autoencoder Framework for Reduced Order Modeling of Turbulent Flow via Strategic Field Decomposition}

\author{Xianglong Li\aff{1}, Zeng Liu\aff{2}, Zhan Wang\aff{3}, Kai Wang\aff{4}, Shunxiang Cao\aff{5} \and Guangyao Wang\aff{1,6}}

\affiliation{\aff{1}Centre for Regional Oceans, Department of Ocean Science and Technology, Institute of Smart City Technologies
, and State Key Laboratory of Internet of Things for Smart City, University of Macau, Macau S.A.R.
\aff{2}School of Naval Architecture and Ocean Engineering, Huazhong University of Science and Technology Institution, Wuhan, China
\aff{3}Institute of Mechanics, Chinese Academy of Sciences, Beijing, China
\aff{4}School of Ocean Engineering and Technology, Sun Yat-Sen University, and Southern Marine Science and Engineering Guangdong Laboratory (Zhuhai), Zhuhai, China
\aff{5}Institute for Ocean Engineering, Shenzhen International Graduate School, Tsinghua University, Shenzhen, China
\aff{6}Zhuhai UM Science and Technology Research Institute, Zhuhai, China
}

\corresau{Guangyao Wang, wanggy@um.edu.mo}

\begin{document}
\maketitle

\begin{abstract}
This study proposes a hybrid reduced-order modeling (ROM) framework for the simulation of turbulent flow. The central idea is to decompose flow dynamics according to their temporal characteristics and predict the resulting components individually. The full field is first divided into a sub-field represented by a limited number of proper orthogonal decomposition (POD) modes (named as POD-retained field) and the corresponding residual sub-field (named as POD-truncated field). A frequency-informed POD strategy identifies the retained modes by considering both modal energy and dominant frequency. The evolution of retained POD coefficients, which feature similar temporal scales, is described using a vector autoregressive (VAR) model. In parallel, the POD-truncated field is compressed into a low-dimensional latent space using a Fourier-neural-operator-based Koopman $\beta$-variational autoencoder (FK-$\beta$-VAE), with the latent variables subsequently predicted by a switching-VAR model. Turbulent statistics of the full field are recovered by combining the contributions from the two components. The framework is assessed using turbulent channel flow at a friction Reynolds number of $110$. The predicted Reynolds-stress components, turbulent kinetic energy (TKE), and dominant wavenumber spectra show good agreement with the reference. Moreover, in comparison with an alternative framework of full-field modeling (i.e., without field decomposition), the proposed framework yields more accurate and robust long-term statistical predictions.
\end{abstract}



\section{Introduction}
\label{sec:introduction}

Reduced-order modeling (ROM) has been widely employed to address the prohibitive computational costs associated with the simulation of high-dimensional flow systems, such as channel flows~\citep{Nakamura2021,Eivazi2021}, flows around bluff bodies~\citep{Hasegawa2020}, and fluid--structure interaction problems~\citep{jiang2024balanced,jiang2026model}. Typically, ROM aims at reproducing the essential flow behavior via a compact set of state variables in a low-dimensional space, while a full-order model (FOM) captures flow dynamics with high fidelity by solving the governing equations involving a high-dimensional state space~\citep{rowley2017model}. Consequently, the overall performance of ROM fundamentally depends on the constructed low-dimensional basis.

In general, ROM consists of two main components, i.e. a method that maps the original high-dimensional onto a low-dimensional state space and a model that describes the dynamics within the latter. The state space mapping methods can be broadly classified into linear and nonlinear types. The linear mapping methods, such as proper orthogonal decomposition (POD) and dynamic mode decomposition (DMD)~\citep{taira2017modal,sirovich1987turbulence,berkooz1993proper}, assume that the high-dimensional data lies on a flat linear subspace and rely on linear transformation operators to realize the space projection. Once the low-dimensional space is established, the corresponding dynamic model can be constructed using either intrusive or non-intrusive techniques. Intrusive methods (e.g., Galerkin projection) derive the low-dimensional equations directly from the original governing equations by forcing the residual to be orthogonal to the subspace spanned by the basis functions and are therefore also referred to as physics-based methods~\citep{wang2012proper,ahmed2021closures}. 
In contrast, non-intrusive methods (e.g. deep learning) treat the time evolution as a black-box function problem without considering the governing equations~\citep{Racca2023,jiang2025koopman}. Instead, they learn the dynamics directly from data and thus are also known as data-driven methods.

However, linear mapping methods often struggle to efficiently represent highly transient or advection-dominated flows, as describing these phenomena on a flat subspace requires an extremely large number of modes~\citep{Maulik2021,Lee2020}. To overcome this limitation, nonlinear mapping methods project the high-dimensional data onto a curved lower-dimensional manifold. Deep learning models, such as autoencoders (AEs) and their probabilistic variants, have become the primary pathway for this task. By utilizing nonlinear activation functions, these models can realize significant dimension compression even for complex dynamics~\citep{Murata2020,Fukami2020Hierarchical,SoleraRico2024}. Because this latent space is constructed through nonlinear transformations, the subsequent temporal modeling is almost exclusively implemented via the aforementioned non-intrusive, data-driven techniques.

Despite the existing ROM methods, which are featured with different low-dimensional representation and dynamic modeling methods, have demonstrated feasibility and success across various problems, almost all of them follow a monolithic strategy. Specifically, they compress and model the entire flow field, which inherently consists of multi-scale dynamics, within a single framework. In other words, the full flow field is compressed directly without decoupling or isolating dynamics with distinct physical scales. This monolithic strategy can potentially result in limited dimensionality reduction, underestimated turbulent statistics, or both, although the specific underlying reasons may vary across different frameworks.
For example, ~\cite{khoo2022sparse} develop a ROM framework for turbulent plane Couette flow by combining the POD-based space mapping method and a data-driven regression model. Their results show that the fist $4$ leading POD modes alone contain approximately $95\%$ of the total perturbation kinetic energy. However, even when the first $41$ POD modes are retained, the ROM still exhibits large deviations from the DNS results, with the velocity fluctuation errors up to $50\%$. Besides,~\cite{halder2026koopman} propose one ROM method by combining the Koopman $\beta$-variational autoencoder and LSTM, which is tested on simulations of flow past a Windsor body. As indicated by their results, the turbulent kinetic energy is underestimated by up to $60\%$. This substantial loss is mainly induced by the nonlinear space mapping, which acts as an overly aggressive dynamics filter by eradicating the relatively small-scale physical processes.

In this regard, we propose a hybrid ROM framework, that explicitly separates the flow dynamics with different temporal characteristics and models their evolutions independently. Specifically, the full field is first decomposed into a subfield represented by a set of retained POD modes (named as POD-retained field) and the corresponding truncated field (named as POD-truncated field). The retained modes are determined using a frequency-informed POD strategy, in which both modal energy and dominant frequency are considered for mode selection. The retained POD coefficients, which are featured with similar temporal characteristics, are then modeled using a vector autoregressive (VAR) model. The POD-truncated field is compressed into a low-dimensional latent space using a Fourier-neural-operator-based Koopman $\beta$-variational autoencoder (FK-$\beta$-VAE), and a switching-VAR model is adopted for the prediction of latent variables.  The proposed framework focuses on the reconstruction of long-term turbulent statistics by combining the contributions from the two components. The method is assessed using a turbulent channel flow at a friction Reynolds number ${\Rey}_{\tau}=110$. The results show that the proposed ROM accurately reproduces the Reynolds stress components, turbulent kinetic energy (TKE) and the wavenumber spectra. In particular, the predicted TKE captures the reference peak location exactly, with only a $3.61\%$ difference in the peak amplitude. In comparison with an alternative framework of direct full-field modeling, which underestimates the Reynolds stress up to $37.27\%$, the proposed method yields an overestimation of only $3.90\%$, corresponding to reduction of $89.54\%$ terms of the absolute error. These results demonstrate that separating flow dynamics according to their temporal characteristics can substantially improve the robustness and accuracy of long-term predictions.

The paper is organized as follows. The proposed hybrid ROM framework and its detailed formulation are introduced in Section~\ref{sec:methodology}. The numerical setup and results for a turbulent channel flow are presented in Section~\ref{sec:numerical_results}. We give the main conclusions in Section~\ref{sec:conclusions}.
\section{Methodology}
\label{sec:methodology}
This section presents the details of the proposed hybrid ROM framework, which aims at predicting the long-term turbulent statistics. In the following subsections, we first introduce the general hybrid ROM framework and then present the sub-modules in detail.
\subsection{The general hybrid ROM framework}
In this study, we consider decomposing the full flow-fluctuation field $\boldsymbol{\xi}$ into two components, one corresponding to the subfield characterized by a limited number of POD modes (named as POD-retained field, $\boldsymbol{\xi^\text{ret}}$) and the other representing the POD truncation residual (named as POD-truncated field, $\boldsymbol{\xi^{\text{tru}}}$). Specifically, we first select the POD modes based on both their energy contributions and frequency features, instead of only considering the former (see \S\ref{subsec:frequency_informed_pod} for details). The retained POD coefficients are modeled via the vector autoregressive (VAR) method (\S\ref{subsec:var_model}).  Furthermore, $\boldsymbol{\xi^{\text{tru}}}$ is mapped into a low-dimensional latent space using a Fourier-neural-operator-based Koopman $\beta$-variational autoencoder (named as FK-$\beta$-VAE), which imposes stronger regularization on the latent variables than conventional autoencoders (\S\ref{subsec:fno_koopman_beta_vae}). The latent variables are then predicted by a switching-VAR method, which combines two VAR predictors with
different lag orders to ensure the deterministic forecasting accuracy across varying time horizons  (see~\S~\ref{subsec:switching_var} for details).  Finally, the turbulent statistics of $\boldsymbol{\xi}$ are obtained by combining the contributions from $\boldsymbol{\xi^\text{ret}}$ and $\boldsymbol{\xi^{\text{tru}}}$.

\begin{figure}
  \centerline{\includegraphics[width=1.1\textwidth]{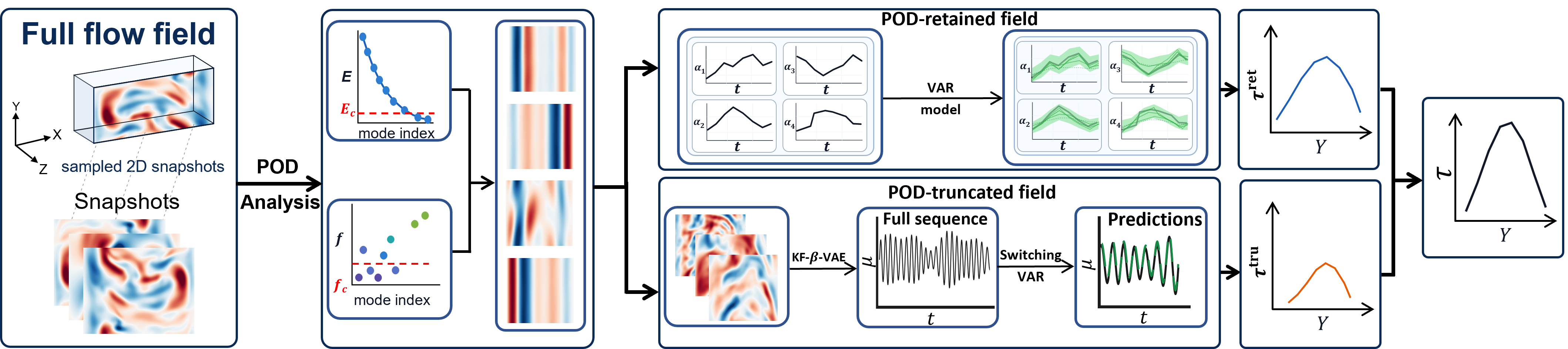}}
  \caption{Hybrid POD-ROM and latent-space truncation-error closure framework. The POD branch models the resolved large-scale component, while the latent residual branch models the unresolved truncation-error statistics.}
\label{fig:hybrid_rom_closure}
\end{figure}

\subsection{Frequency-informed POD}
\label{subsec:frequency_informed_pod}

\subsubsection{POD}
\label{subsubsec:pod_representation}

We consider $N$ snapshots of the fluctuation field,
$\boldsymbol{\xi}_n\in\mathbb{R}^{N_s}$, at time instants $t_n$, $n=1,2,\ldots,N$. $N_s=N_X \times N_Y$ is the total number of
spatial grid points, with $N_X$ and $N_Y$ denoting the numbers of grid
points in the $X$ and $Y$ directions, respectively. Then the snapshot matrix is assembled as
\begin{equation}
\label{eq:snapshot_matrix_pod}
    \boldsymbol{\Xi}
    =
    \left[
    \boldsymbol{\xi}_1,
    \boldsymbol{\xi}_2,
    \ldots,
    \boldsymbol{\xi}_N
    \right]
    \in
    \mathbb{R}^{N_s\times N}.
\end{equation}
Afterwards, the correlation matrix is evaluated as
\begin{equation}
\label{eq:correlation_matrix_pod}
    \boldsymbol{C}
    =
    \boldsymbol{\Xi}^{\text{T}}
    \boldsymbol{\Xi}.
\end{equation}
The eigenvectors $\boldsymbol{\phi}_i$ and eigenvalues $\lambda_i$ are obtained by solving the eigenvalue problem
\begin{equation}
\label{eq:eigenvalue_pod}
    \boldsymbol{C}
    \boldsymbol{\phi}_{i}
    =
    \lambda_{i}
    \boldsymbol{\phi}_{i},
    \qquad
    i=1,2,\ldots,N,
\end{equation}
where the eigenvalues are ordered as
\begin{equation}
    \lambda_{1}
    \ge
    \lambda_{2}
    \ge
    \cdots
    \ge
    \lambda_{N}.
\end{equation}
Then the corresponding orthonormal POD basis vectors are obtained as
\begin{equation}
\label{eq:pod_basis_vectors}
    \boldsymbol{\varphi}_{i}
    =
    \frac{1}{\sqrt{\lambda_{i}}}
    \boldsymbol{\Xi}
    \boldsymbol{\phi}_{i}
    \in
    \mathbb{R}^{N_s\times 1},
    \qquad
    i=1,2,\ldots,N.
\end{equation}
Then the snapshot at the $n$-th time instant can be represented as
\begin{equation}
\label{eq:pod_snapshot_representation}
    \boldsymbol{\xi}_n
    =
    \boldsymbol{\Phi}
    \boldsymbol{a}^{n}
\end{equation}
where
\begin{equation}
\label{eq:pod_basis_matrix}
    \boldsymbol{\Phi}
    =
    \left[
    \boldsymbol{\varphi}_1,
    \boldsymbol{\varphi}_2,
    \ldots,
    \boldsymbol{\varphi}_{N}
    \right]
    \in
    \mathbb{R}^{N_s\times N},
\end{equation}
and
\begin{equation}
\label{eq:pod_coefficient_vector_time}
    \boldsymbol{a}^{n}
    =
    \left[
    \alpha_1^n,
    \alpha_2^n,
    \ldots,
    \alpha_{N}^n
    \right]^\text{T}
    \in
    \mathbb{R}^{N}
\end{equation}
is the POD coefficient vector at the $n$-th time instant. Equivalently, all snapshots can be written in matrix form as
\begin{equation}
\label{eq:pod_matrix_representation}
    \boldsymbol{\Xi}
    =
    \boldsymbol{\Phi}
    \boldsymbol{A},
\end{equation}
where
\begin{equation}
        \boldsymbol{A}
    =
    \left[
    \boldsymbol{a}^{1},
    \boldsymbol{a}^{2},
    \ldots,
    \boldsymbol{a}^{N}
    \right]
    \in
    \mathbb{R}^{N\times N}.
\end{equation}
Moreover, the $i$-th row of $\boldsymbol{A}$,~denoted as~$   \boldsymbol{\alpha}_i=\left[\alpha_i^1,\alpha_i^2,\ldots,\alpha_i^N\right]\in\mathbb{R}^{1\times N}$, represents the temporal sequence of POD coefficients for the $i$-th POD mode.

\subsubsection{Frequency-informed mode selection}
\label{subsubsec:energy_frequency_selection}

The traditional POD truncation method usually selects the retained modes according to their cumulative modal energy ratio, which can be evaluated based on the eigenvalues. In this study, both modal energy and dominant frequency are considered to determine which modes are included to construct $\boldsymbol{\xi}^{\text{ret}}$. Specifically, the relative energy ratio of the $i$-th mode is defined as
\begin{equation}
\label{eq:relative_modal_energy}
    E_i
    =
    \frac{\lambda_i}
    {\sum_{j=1}^{N}\lambda_j},
    \qquad
    i=1,2,\ldots,N.
\end{equation}
The dominant frequency of the $i$-th POD mode is evaluated based on $\boldsymbol{\alpha}_i$. Specifically, the modal spectrum is computed as
\begin{equation}
\label{eq:modal_spectrum}
    \mathcal{P}_i(f)
    =
    \left|
    \mathcal{F}_t
    \left(
    \boldsymbol{\alpha}_i
    \right)
    \right|^2,
\end{equation}
where $\mathcal{F}_t$ denotes the temporal Fourier transform and $f$ denotes the frequency. Then the dominant frequency for the $i$-th mode is determined as
\begin{equation}
    f_i^D
    =
    \operatorname*{arg\,max}_{f}
    \mathcal{P}_i(f).
    \label{eq:dominant_modal_frequency}
\end{equation}
Afterwards, the retained modal index set is formed as
\begin{equation}
\label{eq:frequency_informed_index_set}
    \mathcal{I}
    =
    \left\{
    i
    \ \middle|\
    E_i
    \ge
    E_c,
    \ 
    f_i^D
    \le
    {f_c}
    \right\},
\end{equation}
where $E_c$ and $f_c$ are two predefined criteria for mode selection. In addition, the corresponding subsets of POD basis and coefficients can be formulated as
\begin{equation}
\label{eq:frequency_selected_basis}
    \boldsymbol{\Phi}'
    =\left\{\boldsymbol{\varphi_i}|i\in \mathcal{I} \right\}\in\mathbb{R}^{N_s\times r}.
\end{equation}
and 
\begin{equation}
\label{eq:frequency_selected_coefficients}
    \boldsymbol{A}'=\left\{\boldsymbol{\alpha}_i|i\in\mathcal{I}\right\}\in
    \mathbb{R}^{r\times N}
\end{equation}
where $r$ is the number of retained modes.
Finally, the POD-retained field can be written as
\begin{equation}
\label{eq:frequency_selected_representation}
    \boldsymbol{\Xi}^{\text{ret}}
    =
    \boldsymbol{\Phi}'
    \boldsymbol{A}'.
\end{equation}
The purpose of doing this is to only keep the modes with significant energy levels and comparable frequencies, thereby making it feasible to forecast their corresponding coefficients within a unified framework.

\subsection{VAR Method}
\label{subsec:var_model}

The VAR method is a statistical model of multivariate temporal dynamics~\citep{lutkepohl2005new}. For a state vector
at time $t=t_n$, $\boldsymbol{y}_n\in\mathbb{R}^{d}$, a VAR model with order $q$ is expressed as
\begin{equation}
    \boldsymbol{y}_n
    =
    \boldsymbol{b}
    +
    \sum_{j=1}^{q}
    \boldsymbol{A}_j
    \boldsymbol{y}_{n-j}
    +
    \boldsymbol{\varepsilon}_n,
    \label{eq:var_general}
\end{equation}
where $\boldsymbol{b}\in\mathbb{R}^{d}$ is the intercept vector,
$\boldsymbol{A}_j\in\mathbb{R}^{d\times d}$ is the coefficient matrix for the $j$-step lagged state
$\boldsymbol{y}_{n-j}$, and
$\boldsymbol{\varepsilon}_n\in\mathbb{R}^{d}$ is the error term. For a sequence
$\{\boldsymbol{y}_n\}_{n=1}^{N}$, the target states are collected in a matrix
\begin{equation}
    \boldsymbol{\psi}
    =
    \begin{bmatrix}
        \boldsymbol{y}_{q+1}^{\text{T}} \\
        \boldsymbol{y}_{q+2}^{\text{T}} \\
        \vdots \\
        \boldsymbol{y}_{N}^{\text{T}}
    \end{bmatrix}
    \in
    \mathbb{R}^{(N-q)\times d},
    \label{eq:var_target_matrix}
\end{equation}
and the corresponding lagged states are assembled into the design matrix
\begin{equation}
    \boldsymbol{\Psi}
    =
    \begin{bmatrix}
        1
        & \boldsymbol{y}_{q}^{\text{T}}
        & \boldsymbol{y}_{q-1}^{\text{T}}
        & \cdots
        & \boldsymbol{y}_{1}^{\text{T}}
        \\
        1
        & \boldsymbol{y}_{q+1}^{\text{T}}
        & \boldsymbol{y}_{q}^{\text{T}}
        & \cdots
        & \boldsymbol{y}_{2}^{\text{T}}
        \\
        \vdots
        & \vdots
        & \vdots
        & \ddots
        & \vdots
        \\
        1
        & \boldsymbol{y}_{N-1}^{\text{T}}
        & \boldsymbol{y}_{N-2}^{\text{T}}
        & \cdots
        & \boldsymbol{y}_{N-q}^{\text{T}}
    \end{bmatrix}
    \in
    \mathbb{R}^{(N-q)\times(1+qd)}.
    \label{eq:var_design_matrix}
\end{equation}
The regression parameters are represented in a matrix collectively 

\begin{equation}
    \boldsymbol{B}
        =
    \begin{bmatrix}
        \boldsymbol{b}^{\text{T}} \\
        \boldsymbol{\mathcal{B}}
    \end{bmatrix}
    \in
    \mathbb{R}^{(1+qd)\times d},
    \label{eq:var_regression_matrix}
\end{equation}
where
\begin{equation}
    \boldsymbol{\mathcal{B}}
    =
    \begin{bmatrix}
        \boldsymbol{A}_1^{\text{T}} \\
        \boldsymbol{A}_2^{\text{T}} \\
        \vdots \\
        \boldsymbol{A}_q^{\text{T}}
    \end{bmatrix}
    \in
    \mathbb{R}^{(qd)\times d}
    \label{eq:var_coefficient_matrix}
\end{equation}
contains the lag coefficient matrices.

Then the system of VAR equations corresponding to different time instants can be written in a matrix form as
\begin{equation}
    \boldsymbol{\psi}
    =
    \boldsymbol{\Psi}\boldsymbol{B}
    +
    \boldsymbol{\mathcal{E}},
    \label{eq:var_matrix_form}
\end{equation}
where
\begin{equation}
    \boldsymbol{\mathcal{E}}
    =
    \begin{bmatrix}
        \boldsymbol{\varepsilon}_{q+1}^{\text{T}} \\
        \boldsymbol{\varepsilon}_{q+2}^{\text{T}} \\
        \vdots \\
        \boldsymbol{\varepsilon}_{N}^{\text{T}}
    \end{bmatrix}
    \in
    \mathbb{R}^{(N-q)\times d}
    \label{eq:var_error_matrix}
\end{equation}
is the error matrix.

Afterwards, the optimal regression parameters are obtained by solving a
ridge-regularized least-squares problem~\citep{hoerl1970ridge},
\begin{equation}
    \boldsymbol{B}^{\mathrm{opt}}
    =
    \arg\min_{\boldsymbol{B}}
    \left(
        \left\|
        \boldsymbol{\psi}
        -
        \boldsymbol{\Psi}\boldsymbol{B}
        \right\|_{F}^{2}
        +
        \eta
        \left\|
        \boldsymbol{\mathcal{B}}
        \right\|_{F}^{2}
    \right),
    \label{eq:var_ridge}
\end{equation}
where $\boldsymbol{B}^{\mathrm{opt}}$ denotes the optimal
ridge-regression coefficient matrix, $\|\cdot\|_{F}$ denotes the
Frobenius norm, and $\eta\geq0$ controls the regularization strength. Compared with
conventional least squares, ridge regression introduces an additional
quadratic penalty on the lag-coefficient matrix, reducing the variance
of the estimated coefficients at the expense of a small bias and thereby
improving model robustness. The corresponding  residual matrix is evaluated as
\begin{equation}
    \boldsymbol{\mathcal{E}}
    =
    \boldsymbol{\psi}
    -
    \boldsymbol{\Psi}\boldsymbol{B}^{\mathrm{opt}},
    \label{eq:var_fitted_error_matrix}
\end{equation}
The residual covariance matrix is then calculated as
\begin{equation}
    \boldsymbol{\Omega}
    =
    \frac{\boldsymbol{\mathcal{E}}^{\text{T}} \boldsymbol{\mathcal{E}}}{N-q-1}.
    \label{eq:var_residual_covariance}
\end{equation}
Since this study aims at predicting the turbulent statistics of the flow
field, Monte Carlo simulations are subsequently performed. For the same
prediction horizon, $M$ trajectories are generated recursively using
Eq.~\eqref{eq:var_general} with the fitted regression parameters, while
the innovation $\boldsymbol{\varepsilon}_n$ at each time step is sampled
from
$\mathcal{N}(\boldsymbol{0},\boldsymbol{\Omega})$. Finally, the contribution from the retained POD modes to the turbulent statistics is evaluated based on the reconstructed snapshots with VAR-predicted POD coefficients.

\subsection{Fourier-neural-operator-based Koopman $\beta$-variational autoencoder}
\label{subsec:fno_koopman_beta_vae}

\begin{figure}
  \centerline{\includegraphics[width=0.95\textwidth]{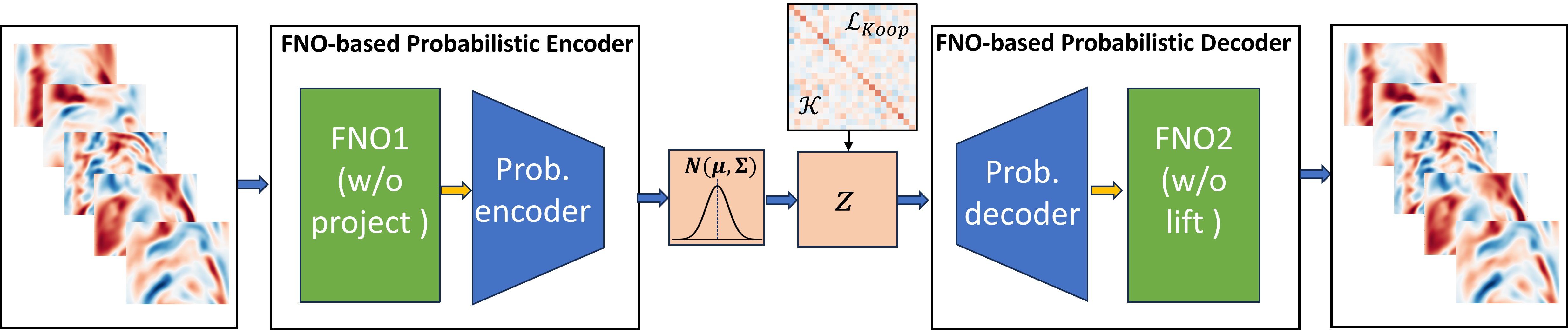}}
  \caption{Schematic of the FK-\(\beta\)-VAE. The backbone is an FNO-enhanced \(\beta\)-VAE for residual-field representation. The FNO operator performs spectral--pointwise spatial feature mixing, while the Koopman branch regularizes the latent-mean evolution during training.}
  \label{fig:koopman_fno_beta_vae_overview}
\end{figure}

The POD-truncated field is compressed and represented by a Fourier-neural-operator-based Koopman $\beta$-variational autoencoder, for which the overall structure is shown in  figure~\ref{fig:koopman_fno_beta_vae_overview}. This model is built on the basis of VAE, which maps each input field to a probability distribution in the latent space and then reconstructs the field from the sampled latent variable. Practically, we adopt $\beta$-VAE for the trade-off between reconstruction accuracy and interpretability. Moreover, Fourier neural operators (FNOs) are integrated into the model as spatial representation modules to capture both local and nonlocal correlations. Additionally, a linear Koopman transition is imposed on the latent mean space during training, so that the encoded temporal sequence is regularized for subsequent latent-space forecasting.

\subsubsection{$\beta$-Variational Autoencoder}
\label{subsubsec:beta_vae}

The VAE first maps $\boldsymbol{\xi^\text{tru}}$ (the time index is dropped off in this section for the purpose of conciseness) through convolutional layers into a normal distribution within a latent space, $\boldsymbol{z}\sim\mathcal{N}\left(\boldsymbol{\mu},\boldsymbol{\Sigma}\right)$,
where $\boldsymbol{z}\in\mathbb{R}^{d_z}$ is the $d_z$-dimensional latent variable, and $\boldsymbol{\mu}$ and
$\boldsymbol{\Sigma}$ are its mean and diagonal covariance matrix, respectively. In practice, the encoder outputs the mean $\boldsymbol{\mu}$ and the logarithm of variance terms $l_{i}=\log\left[\boldsymbol{\Sigma}(i,i)\right],~i=1,2\cdots d_z$, to guarantee positive variance values and enhance numerical stability. Afterwards, the latent variable is generated via the reparameterization trick,
\begin{equation}
    \boldsymbol{z}
    =
    \boldsymbol{\mu}
    +
    \exp
    \left(
        \frac{1}{2}
        \boldsymbol{\ell}
    \right)
    \odot
    \boldsymbol{\epsilon},
    \qquad
    \boldsymbol{\epsilon}
    \sim
    \mathcal{N}
    \left(
        \boldsymbol{0},
        \boldsymbol{I}
    \right),
    \label{eq:vae_reparameterization}
\end{equation}
where $\boldsymbol{\ell}=[l_1,l_2,\cdots,l_{d_z}]^{\text{T}}$ and $\odot$ denotes element-wise multiplication. The decoder then builds the reconstructed residual snapshot $\widehat{\boldsymbol{\xi}}^\text{tru}$ from the sampled latent variable.

The loss of VAE consists of the reconstruction and latent distribution regularization loss terms. The reconstruction loss term measures the discrepancy between the decoded and original residual snapshots, which is defined as
\begin{equation}
    \mathcal{L}_{\mathrm{rec}}
    =||\widehat{\boldsymbol{\xi}}^\text{tru}-\boldsymbol{\xi}^\text{tru}||_F^2
    \label{eq:vae_rec_loss}
\end{equation}
The latent distribution regularization term is defined as the KL divergence between the approximate posterior $\mathcal{N}\left(\boldsymbol{\mu},\boldsymbol{\Sigma}\right)$ and the standard normal prior $p(\boldsymbol{z})=\mathcal{N}(\boldsymbol{0},\boldsymbol{I})$, which is formulated as
\begin{equation}
    \mathcal{L}_{\mathrm{KL}}
    =
    -\frac{1}{2}
    \sum_{i=1}^{d_z}
    \left[
        1
        +
        l_{i}
        -
        \left(
            \mu_{i}
        \right)^2
        -
        \exp
        \left(
            l_{i}
        \right)
    \right].
    \label{eq:vae_kl_loss}
\end{equation}
As a variant of the conventional VAE, $\beta$-VAE introduces one additional hyperparameter $\beta$ 
to control the tradeoff between $\mathcal{L}_{\mathrm{rec}}$ and $\mathcal{L}_{\mathrm{KL}}$, for which the corresponding loss term is defined as
\begin{equation}
    \mathcal{L}_{\beta\text{-VAE}}
    =
    \mathcal{L}_{\mathrm{rec}}
    +
    \beta
    \mathcal{L}_{\mathrm{KL}}.
    \label{eq:beta_vae_loss}
\end{equation}
The parameter $\beta$ therefore controls the balance between reconstruction accuracy and latent-space regularization.

\subsubsection{Fourier neural operator}
\label{subsubsec:fno_blocks}

The FNO is integrated into the $\beta$-VAE to improve the spatial representation of residual fields. Although the POD-truncated field is obtained by removing the high-energy POD contributions from the original field, it still contains considerable spatially correlated structures. A conventional encoder captures nonlocal dependencies only through the progressive expansion of the receptive field. By contrast, the FNO performs global spectral mixing over the sampled grid, which allows the low-wavenumber, nonlocal structures to be represented directly via a finite set of retained Fourier modes.

Within the FNO module, the single-channel input field is first lifted to a multi-channel feature field with the same spatial resolution,
\begin{equation}
    \boldsymbol{h}^{(1)}
    =
    \mathfrak{L}
    \left(
        \boldsymbol{\xi^\text{tru}}
    \right)
    \in
    \mathbb{R}^{C_1\times N_X\times N_Y},
    \label{eq:fno_lifting}
\end{equation}
where $\mathfrak{L}$ denotes the lifting operator, and $C_1$ is the channel dimension. Then the lifted feature field passes through $M$ stacked Fourier layers. For the $m$-th FNO layer, the feature field is updated as 
\begin{equation}
    \boldsymbol{h}^{(m+1)}
    =
    \varrho_{\mathrm{GELU}}
    \left\{
    \operatorname{BN}
    \left[
    \mathcal{F}^{-1}
    \left(
        \mathcal{R}^{(m)}
        \left(
            \mathcal{F}
            \left[
                \boldsymbol{h}^{(m)}
            \right]
        \right)
    \right)
    +
    \mathcal{W}_{\mathrm{p}}^{(m)}
    \boldsymbol{h}^{(m)}
    \right]
    \right\}.
    \label{eq:fno_block_compact}
\end{equation}
where $\mathcal{F}$ and $\mathcal{F}^{-1}$ denote the fast Fourier transform (FFT) and inverse FFT, respectively. $\mathcal{R}^{(m)}$ is the learnable complex-valued tensor representing the kernel in the frequency domain. $\mathcal{W}_{\mathrm{p}}^{(m)}$ is a point-wise linear transformation, which is applied on $\boldsymbol{h}^{(m)}$ in order to capture the  high-frequency local features that might be lost during frequency truncation. $\operatorname{BN}$ and $\varrho_{\mathrm{GELU}}$ denote batch normalization and GELU activation function, respectively. Finally, after $M$ Fourier layers, a projection layer maps $\boldsymbol{h}^{(M+1)}$ back to a single-channel field,
\begin{equation}
    \widehat{\boldsymbol{\xi}}^\text{tru}
    =
    \mathfrak{P}
    \left(
        \boldsymbol{h}^{(M+1)}
    \right)\in
    \mathbb{R}^{1\times N_X\times N_Y},
    \label{eq:fno_output_projection}
\end{equation}
where $\mathfrak{P}$ is the projection operator.

\subsubsection{Koopman Latent Regularization}
\label{subsubsec:koopman_loss_function}

The Koopman regularization is introduced to  encourage linear temporal evolution of the latent state, thereby rendering subsequent forecasting more tractable. The latent mean is used as the constrained state $\boldsymbol{s}_n$. Then a trainable Koopman operator $\mathcal{K}\in   \mathbb{R}^{d_z\times d_z}$  is introduced as
\begin{equation}
    \boldsymbol{s}_{n+1}
    =
    \mathcal{K}
    \boldsymbol{s}_{n}.
    \label{eq:koopman_map}
\end{equation}
For a training window $\{t_n,t_{n+1},\ldots,t_{n+T}\}$, the Koopman rollout is initialized from the reference state of the first time instant, $\boldsymbol{s}_{n}$, and the approximated sequence $\left\{\tilde{\boldsymbol{s}}_{n+1},\tilde{\boldsymbol{s}}_{n+2},\cdots\tilde{\boldsymbol{s}}_{n+T}\right\}$ is obtained by applying $\mathcal{K}$ recursively. The Koopman loss is defined as
\begin{equation}    
    \mathcal{L}_{\mathrm{Koop}} =\mathcal{L}_\text{lin}+\zeta_\text{pred}\mathcal{L}_\text{pred}+\zeta_\text{orth}\mathcal{L}_\text{orth},
\end{equation}
where 
\begin{equation}
\mathcal{L}_\text{lin}=\frac{1}{T}
    \sum_{k=1}^{T}
    \left\|
        \tilde{\boldsymbol{s}}_{n+k}
        -
        \boldsymbol{s}_{n+k}
    \right\|_{2}^{2}
\end{equation}
is the Koopman linear evolution loss. $\mathcal{L}_\text{pred}$ is the prediction loss and formulated as
\begin{equation}
    \mathcal{L}_\text{pred}=\frac{1}{T}
    \sum_{i=1}^{T} \left\|
        \mathcal{D}_{\mathrm{FNO}}
        \left(
            \tilde{\boldsymbol{s}}_{n+i}
        \right)
        -
        \boldsymbol{\xi}^\text{tru}_{n+i}
    \right\|_{F}^{2},
\end{equation}
where $\mathcal{D}_{\mathrm{FNO}}$ denotes the FNO-based probabilistic decoder operator. $\zeta_\text{pred}$ is the corresponding weight. $\mathcal{L}_{\mathrm{orth}}$ is an orthogonality regularization term, which is expressed as
\begin{equation}
    \mathcal{L}_{\mathrm{orth}}
    =
    \left\|
        \mathcal{K}^{T}
        \mathcal{K}
        -
        \boldsymbol{I}
    \right\|_{F},
    \label{eq:koopman_orth_loss}
\end{equation}
This term discourages excessive amplification or decay during recursive application of $\mathcal{K}$ and $\zeta_{\mathrm{orth}}$ is its weight.
Finally, the total loss for the proposed FK-$\beta$-VAE is formulated as
\begin{eqnarray}
    \mathcal{L}
    =
    {\mathcal{L}}_{\beta-\text{VAE}}
    +
    \gamma
    \mathcal{L}_{\mathrm{Koop}},
    \label{eq:koopman_beta_vae_total_loss}
\end{eqnarray}
where $\gamma$ is a parameter which controls the contribution of the Koopman regularization.

During the initial training stage, the model parameters are first optimized with $\gamma=0$, such that we first build a preliminary encoder-decoder architecture without regularizing the latent dynamics. That is to say, only the model parameters in the encoder and decoder operators are updated during this stage. Then the joint training is performed, during which the Koopman weight is linearly increased from zero to its prescribed value during the initial warm-up epochs and is then kept fixed. In this process, the model parameters for both the encoder-decoder architecture and $\mathcal{K}$ are updated simultaneously.  After training, the sequence of latent means is extracted and used as the dataset for the subsequent latent-space forecasting module.

\subsection{Switching-VAR Forecasting}
\label{subsec:switching_var}
Following the latent-space representation described above, temporal
forecasting is performed directly for the latent mean sequence
$\boldsymbol{\mu}_n\in\mathbb{R}^{d_z}$. Here we use the switching-VAR, combining two VAR models with distinct lag orders to enhance the prediction robustness across multiple lead times. Specifically, two VAR models with orders $q^{(1)}$ and $q^{(2)}$ ($q^{(1)}<q^{(2)}$) are fitted to the latent-mean sequence using the procedure described in \S~\ref{subsec:var_model}. Then, motivated by smooth-transition autoregressive modeling~\citep{terasvirta1994star,vanDijk2002STAR}, the two predicted trajectories are combined as
\begin{equation}
    \widehat{\boldsymbol{\mu}}_n
    =
    \left(1-\chi_{n}\right)
    \widehat{\boldsymbol{\mu}}_n^{(1)}
    +
    \chi_{n}
    \widehat{\boldsymbol{\mu}}_n^{(2)},
    \label{eq:var_switching_prediction}
\end{equation}
where $\chi_{n}$ is a raised-cosine switching function.
\begin{equation}
    \chi_n
    =
    \begin{cases}
    0,
    &
    t_n \leq c_s-\Delta_s/2,
    \\[2mm]
    \dfrac{1}{2}
    -
    \dfrac{1}{2}
    \cos
    \left[
        \upi
        \dfrac{
        t_n-\left(c_s-\Delta_s/2\right)
        }{\Delta_s}
    \right],
    &
    c_s-\Delta_s/2
    <
    t_n
    <
    c_s+\Delta_s/2,
    \\[3mm]
    1,
    &
    t_n \geq c_s+\Delta_s/2,
    \end{cases}
    \label{eq:var_switching_weight}
\end{equation}
where $c_s$ and $\Delta_s$ denote the centre and width of the switching interval, respectively. As a result, the early prediction stage is primarily governed by the lower-order predictor; as the horizon advances, the secondary predictor is introduced via a smooth transition weighting and ultimately takes full control of the temporal propagation.

\section{Numerical Experiments}
\label{sec:numerical_results}
This section evaluates the proposed hybrid ROM framework based on the simulation of a turbulent channel flow. Specifically, we first run a DNS model to generate the high-fidelity dataset, which then works as the basis for constructing and validating the hybrid ROM framework.

\subsection{Turbulent Channel-Flow Dataset}
\label{subsec:u_case_data_preparation}

Here we consider an incompressible turbulent channel flow, which is governed by the incompressible Navier--Stokes equations,
\begin{equation}
        \nabla\cdot\boldsymbol{u}=0
    \label{eq:con}
\end{equation}

\begin{equation}
        \frac{\partial\boldsymbol{u}}{\partial t}
        +
        \left(
            \boldsymbol{u}\cdot\nabla
        \right)
        \boldsymbol{u}
        =
        -\nabla p
        +
        \nu\nabla^2\boldsymbol{u}
    \label{eq:channel_governing_equations}
\end{equation}
where $\boldsymbol{u}=(u_1,u_2,u_3)^{\mathrm{T}}$ is the velocity vector, with $u_1,~u_2,~\text{and}~u_3$ denoting the components along $X$, $Y$, and $Z$ directions, respectively. $p$ is the kinematic pressure, and $\nu$ is the kinematic viscosity. The flow condition is characterized by the friction Reynolds number based on the channel half-height,
\begin{equation}
    \Rey_{\tau}
    =
    \frac{u_{\tau}h}{\nu}
    =
    110,
    \label{eq:channel_friction_reynolds_number}
\end{equation}
where $u_{\tau}$ is the friction velocity and $h$ is the channel half-height.
The dataset is obtained from an open-source DNS model,~i.e. Canonical Navier-Stokes (CaNS), developed by Costa~\cite{costa2018cans}. In CaNS, the governing equations are discretized on a staggered Cartesian grid using the second-order finite difference scheme, and the temporal integration is performed using a three-stage Runge--Kutta scheme. Moreover, the pressure Poisson equation is solved using eigenfunction expansions and fast Fourier transforms in the homogeneous directions.

\begin{figure}
  \centerline{\includegraphics[width=0.8\textwidth]{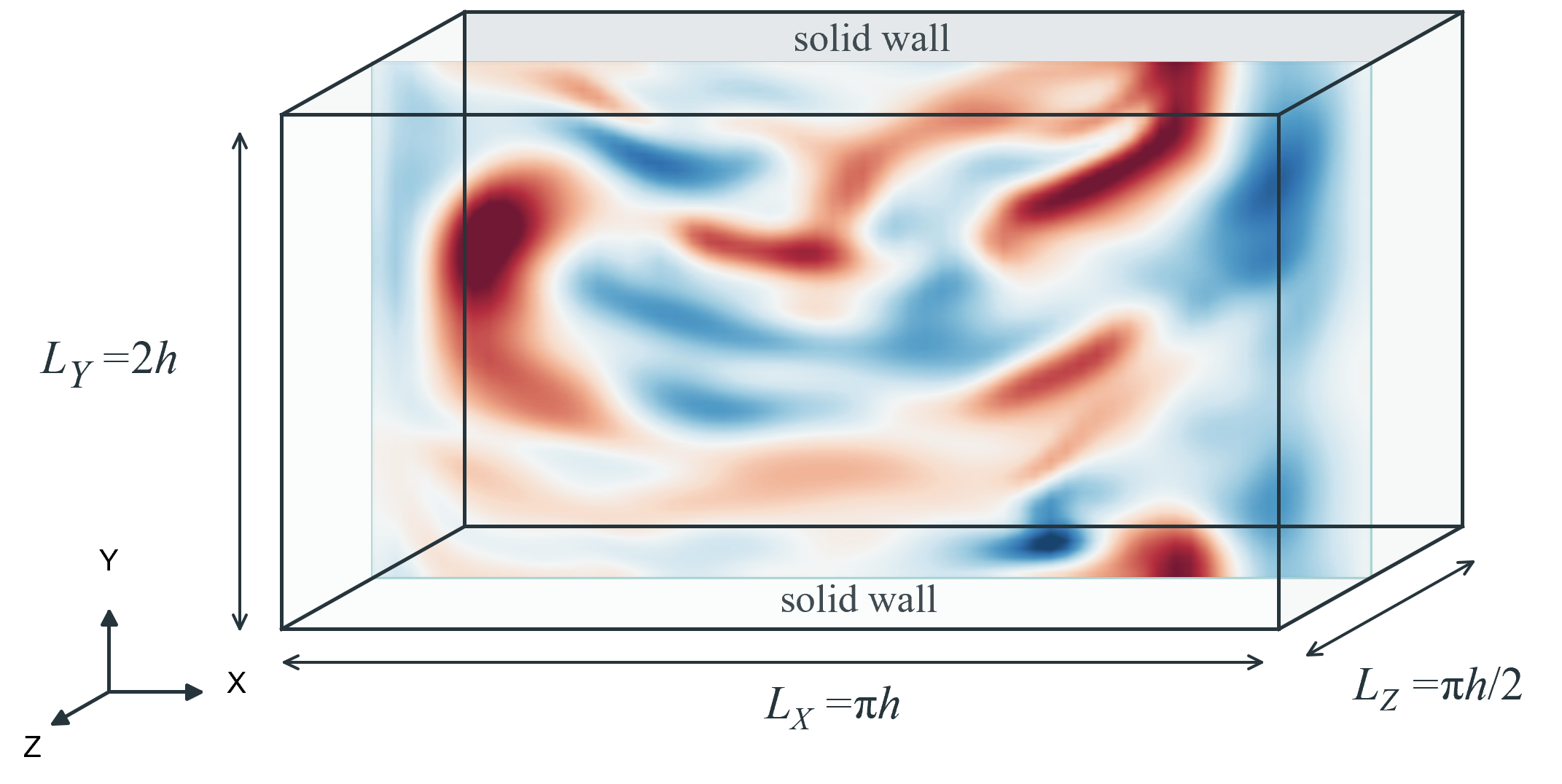}}
  \caption{Computational domain and spanwise mid-plane sampling location for the turbulent channel-flow dataset.}
  \label{fig:u_channel_midplane_sampling}
\end{figure}

The simulation is performed with a time step ${\Delta} t=0.01h/u_{\tau}$. The computational domain sizes are $L_X=\upi h$, $L_Y=2h$ and $L_Z=\upi h/2$, discretized using $64\times64\times64$ grid points. A schematic of the computational
domain is shown in figure~\ref{fig:u_channel_midplane_sampling}. Periodic boundary conditions are imposed in the $X$ and $Z$ directions, whereas no-slip conditions are prescribed at the walls located at $Y=0$ and $Y=2h$. The grid is uniform in the $X$ and $Z$ directions and stretched in the $Y$ direction to improve the near-wall resolution. The flow is initialized from a laminar Poiseuille profile and advanced under a prescribed constant pressure gradient until a statistically stationary turbulent state is established. Subsequently, $12\,000$ two-dimensional snapshots are extracted at $Z=\upi h/4$, with a temporal interval of $10\Delta t$.

\subsection{Frequency-informed POD results}
\label{subsec:u_pod_resolved_results}

\begin{figure}
  \centerline{\includegraphics[width=0.78\textwidth]
  {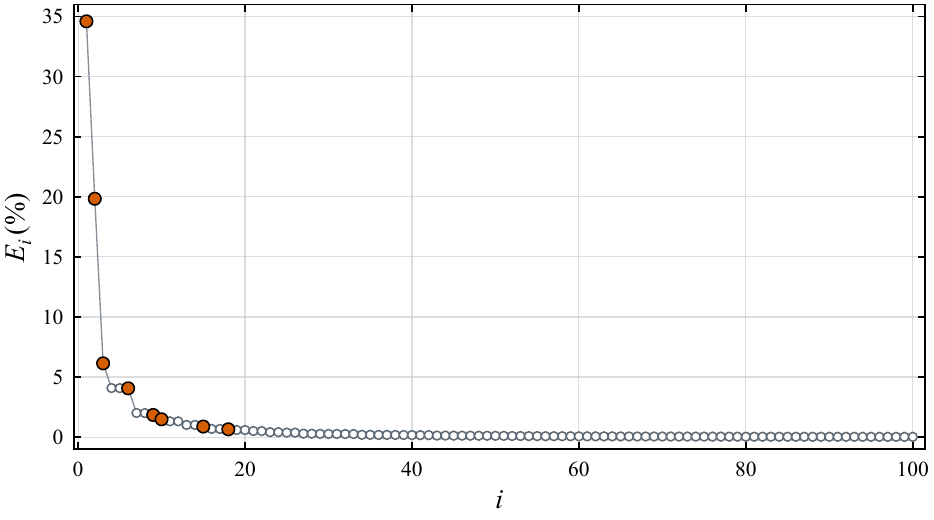}}
  \caption{Modal energy distribution of the first 100 POD modes, with retained modes indicated by filled orange circles and unretained modes by grey circles.}
  \label{fig:u_pod_energy}
\end{figure}

\begin{figure}
  \centerline{\includegraphics[width=0.78\textwidth]
  {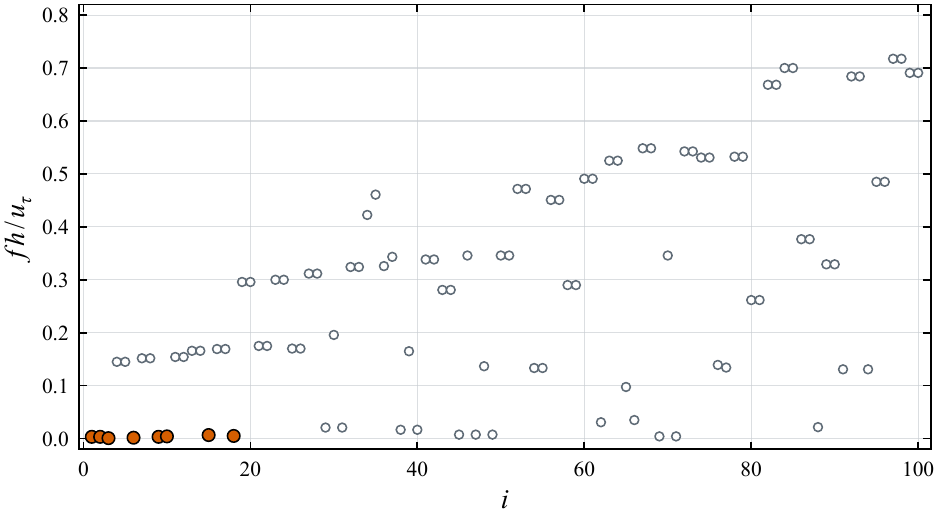}}
  \caption{Dominant-frequency distribution of the first 100 POD modes of the streamwise velocity fluctuation field, with retained modes indicated by filled orange markers and unretained modes by open grey markers.}
  \label{fig:u_pod_frequency}
\end{figure}

The frequency-informed POD is first applied to the snapshots to determine the retained modes, based on which the whole flow field is sequentially divided into the POD-retained and POD-truncated fields, i.e. $\boldsymbol{\xi}^{\text{ret}}$ and $\boldsymbol{\xi}^{\text{tru}}$. Figures~\ref{fig:u_pod_energy} and~\ref{fig:u_pod_frequency} show the relative modal energy ratios $E_i$ and dominant frequencies $f_i^D$ of the first $100$ POD modes, respectively. Specifically, the prescribed criteria for $E_i$ and $f_i^D$ are
${E_c}=0.3\,\%$ and ${f^D_c}=~{0.01 (u_\tau/h)}$, respectively, yielding the retained modal index set
\begin{equation}
    \mathcal{I}
    =
    \left\{
        1,2,3,6,9,10,15,18
    \right\}.
    \label{eq:u_retained_pod_modes}
\end{equation}
The $8$ retained POD modes account for approximately $69.51\,\%$ of the total energy. Moreover, they are featured with comparable dominant frequencies with a mean of $3.54\times10^{-3}(u_\tau/h)$ and variance of only $2.91\times10^{-6}(u_\tau/h)^2$, which therefore makes it feasible to predict their corresponding POD coefficients with a single VAR model.

\begin{figure}
    \centering
    \captionsetup{style=capcenter}

    \begin{subfigure}{0.8\textwidth}
        \centering
        \includegraphics[width=\linewidth]
        {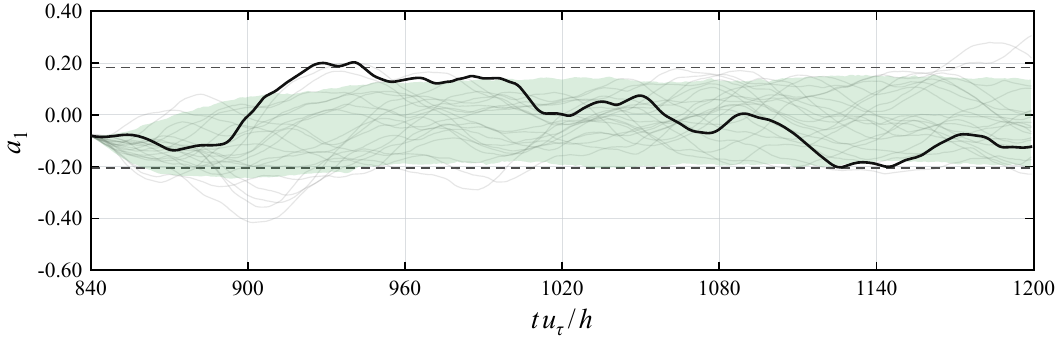}
        \caption{}
        \label{fig:u_pod_coeff_mode01}
    \end{subfigure}

    \vspace{0.4em}

    \begin{subfigure}{0.8\textwidth}
        \centering
        \includegraphics[width=\linewidth]
        {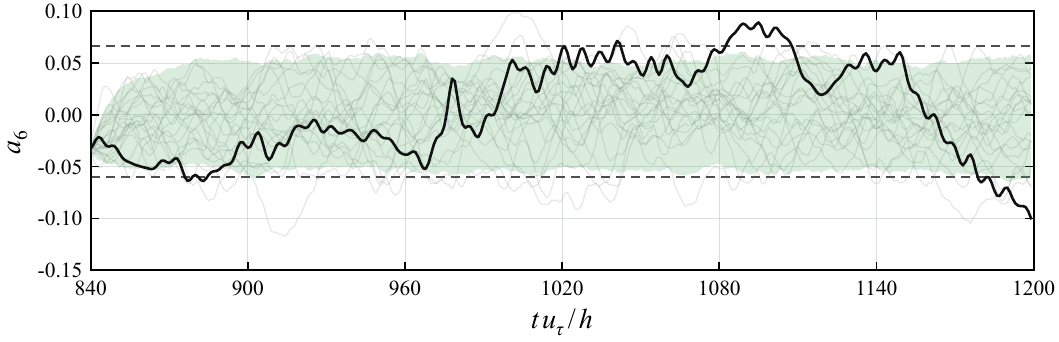}
        \caption{}
        \label{fig:u_pod_coeff_mode06}
    \end{subfigure}

    \vspace{0.4em}

    \begin{subfigure}{0.8\textwidth}
        \centering
        \includegraphics[width=\linewidth]
        {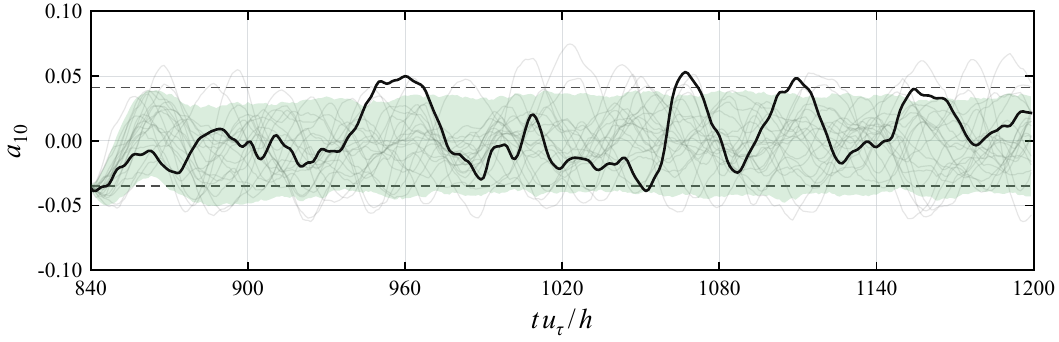}
        \caption{}
        \label{fig:u_pod_coeff_mode10}
    \end{subfigure}

    \vspace{0.4em}

    \begin{subfigure}{0.8\textwidth}
        \centering
        \includegraphics[width=\linewidth]
        {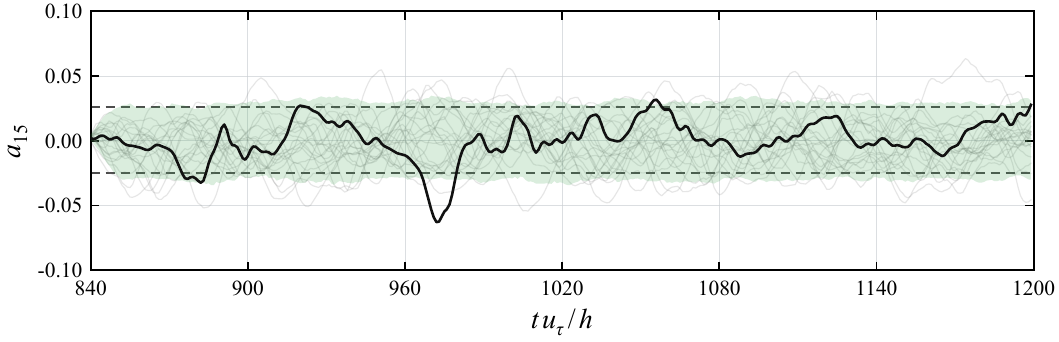}
        \caption{}
        \label{fig:u_pod_coeff_mode15}
    \end{subfigure}

    \caption{Temporal histories of selected POD coefficients from the reference dataset (black solid line) and VAR predictions (grey lines). The reference and predicted 5th--95th percentile ranges are denoted by grey dashed lines and the green shaded area, respectively: (a) $\alpha_1$; (b) $\alpha_6$; (c) $\alpha_{10}$; (d) $\alpha_{15}$.}
    \label{fig:u_pod_coeff_time_histories}
\end{figure}

\begin{figure}
    \centering
    \captionsetup{style=capcenter}

    \begin{subfigure}[t]{0.48\textwidth}
        \centering
        \includegraphics[width=\linewidth]
        {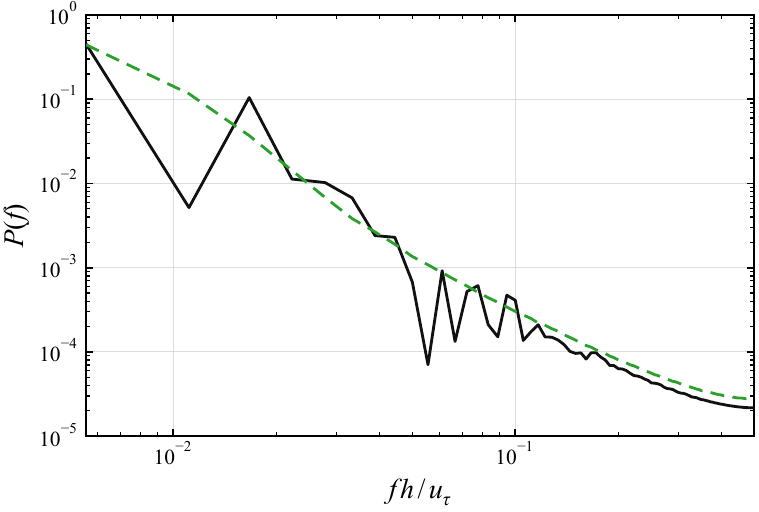}
        \caption{}
        \label{fig:u_pod_spectrum_mode01}
    \end{subfigure}
    \hfill
    \begin{subfigure}[t]{0.48\textwidth}
        \centering
        \includegraphics[width=\linewidth]
        {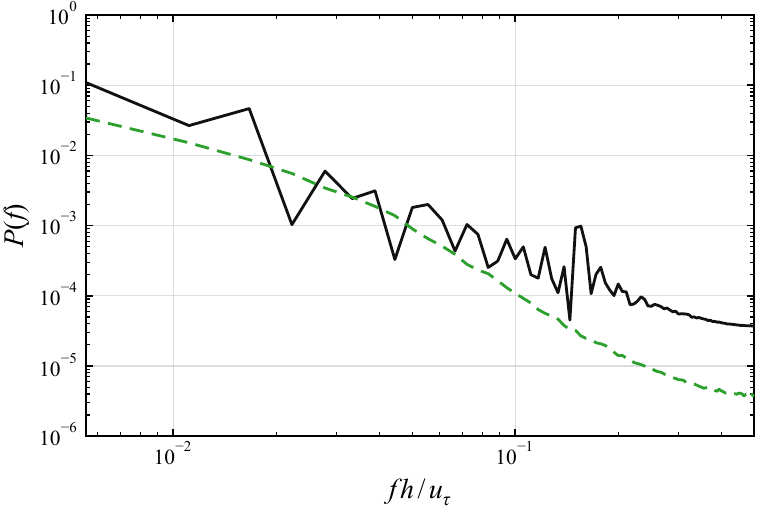}
        \caption{}
        \label{fig:u_pod_spectrum_mode06}
    \end{subfigure}

    \vspace{0.6em}

    \begin{subfigure}[t]{0.48\textwidth}
        \centering
        \includegraphics[width=\linewidth]
        {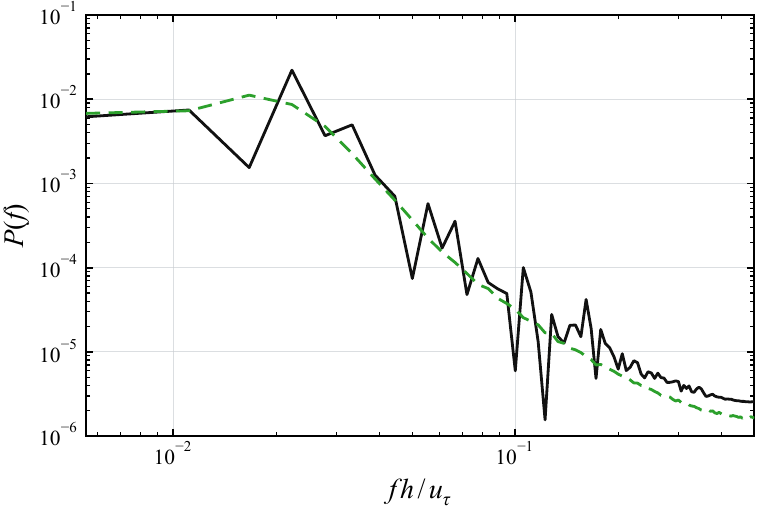}
        \caption{}
        \label{fig:u_pod_spectrum_mode10}
    \end{subfigure}
    \hfill
    \begin{subfigure}[t]{0.48\textwidth}
        \centering
        \includegraphics[width=\linewidth]
        {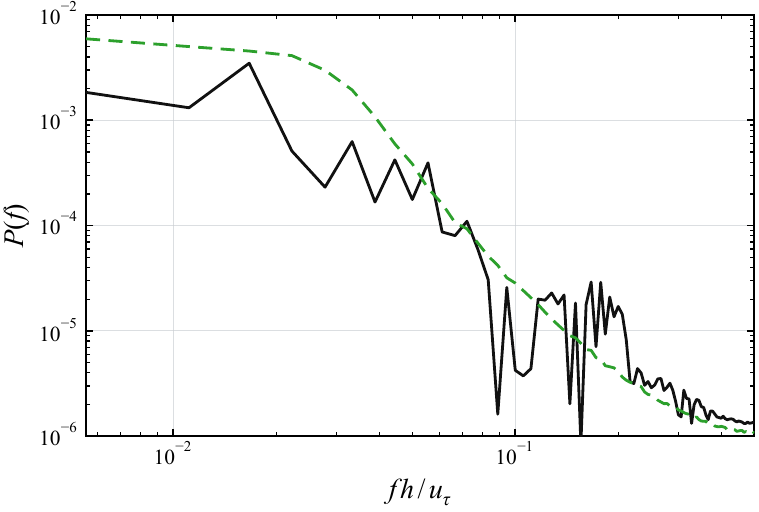}
        \caption{}
        \label{fig:u_pod_spectrum_mode15}
    \end{subfigure}

    \caption{Modal power spectra of selected retained POD coefficients
    from the reference (black solid line) and VAR predictions
    (green dashed line): (a) $\alpha_1$; (b) $\alpha_6$;
    (c) $\alpha_{10}$; (d) $\alpha_{15}$.}
    \label{fig:u_pod_coeff_spectra}
\end{figure}

To build the VAR model, we set $q=2$ and $\eta=10^{-4}$. 
The resulting standardized innovation covariance matrix for the
streamwise component is
\begin{equation}
\boldsymbol{\Omega}
=
10^{-4}
\begin{bmatrix}
  1.29 &  0.06 & -0.20 & -0.40 & -0.95 &  0.21 &  0.08 & -0.79 \\
  0.06 &  2.21 &  0.50 &  0.20 &  0.40 &  0.85 &  2.16 & -0.04 \\
 -0.20 &  0.50 &  5.03 &  0.55 & -6.73 &  0.88 &  1.21 &  0.75 \\
 -0.40 &  0.20 &  0.55 & 66.71 &  0.13 &  2.49 & -3.02 & -14.37 \\
 -0.95 &  0.40 & -6.73 &  0.13 & 81.46 & -9.89 & -8.23 & 10.53 \\
  0.21 &  0.85 &  0.88 &  2.49 & -9.89 & 27.77 &  4.30 & -6.91 \\
  0.08 &  2.16 &  1.21 & -3.02 & -8.23 &  4.30 & 50.92 & 38.80 \\
 -0.79 & -0.04 &  0.75 & -14.37 & 10.53 & -6.91 & 38.80 & 344.09
\end{bmatrix}.
\label{eq:u_var_innovation_covariance}
\end{equation}
To implement the prediction step, $\mathcal{M}=512$ independent trajectories are generated, with the error term $\boldsymbol{\varepsilon}_n$ randomly sampled from $\mathcal{N}(\boldsymbol{0},\boldsymbol{\Omega})$ at each time instant. Figure~\ref{fig:u_pod_coeff_time_histories} presents the predicted POD coefficients for $4$ selected modes (including modes $1,~6,~10,~\text{and}~15$) and their reference, as well as their variance information (presented in the form of the 5th-95th percentile range). It can be observed that the predicted variances demonstrate an initial trend of progressive expansion and eventually converge to constant values. The reason is that
all the predicted trajectories are initialized from the same (reference) POD coefficients, which makes the initial ensemble represent a conditional distribution with relatively small spreads. As the independent random errors are introduced continuously, the influence of the initial condition decays and the predicted variance approaches a stationary value~\cite{lutkepohl2005new}. In this regard, the last $1800$ snapshots are used for further evaluation, where the variances are steady and also consistent with those estimated from the reference trajectories.  Figure~\ref{fig:u_pod_coeff_spectra} compares the power spectra of representative retained POD coefficients obtained from the reference data and the VAR predictions. It can be observed that the VAR model successfully captures the overall spectral feature, especially the spectral decay, of the POD-retained field. In particular, the energetically dominant low-frequency range ($f<0.5 u_\tau/h$), which contains more than $96.4\%$ of the total spectral energy, is accurately reproduced by the VAR predictions. The discrepancies are mainly confined to higher frequencies, where the corresponding energy levels are substantially lower.

\begin{figure}
    \centering
    \captionsetup{style=capcenter}

    \begin{subfigure}{0.75\textwidth}
        \centering
        \includegraphics[width=\linewidth]
        {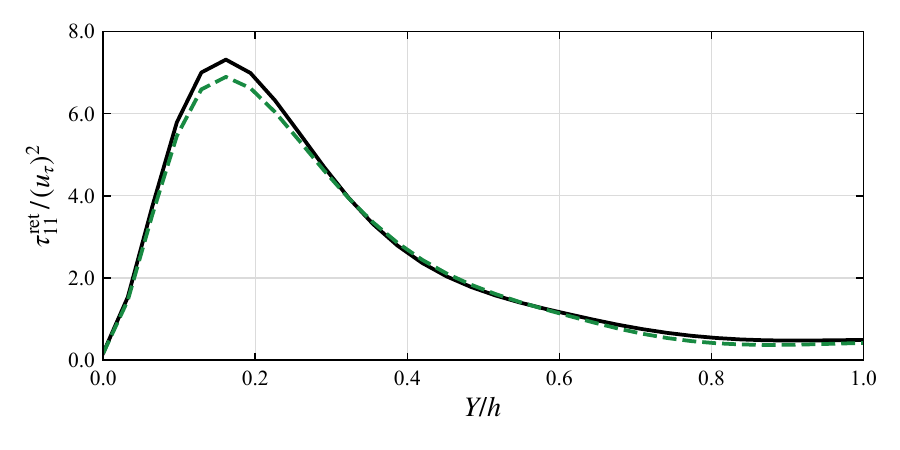}
        \caption{}
        \label{fig:pod_var_rs_u}
    \end{subfigure}

    \vspace{0.1em}

    \begin{subfigure}{0.75\textwidth}
        \centering
        \includegraphics[width=\linewidth]
        {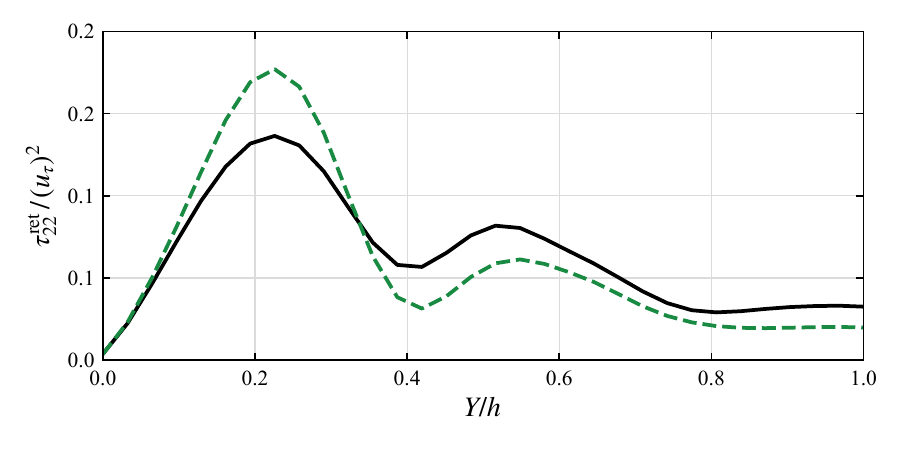}
        \caption{}
        \label{fig:pod_var_rs_v}
    \end{subfigure}

    \vspace{0.1em}

    \begin{subfigure}{0.75\textwidth}
        \centering
        \includegraphics[width=\linewidth]
        {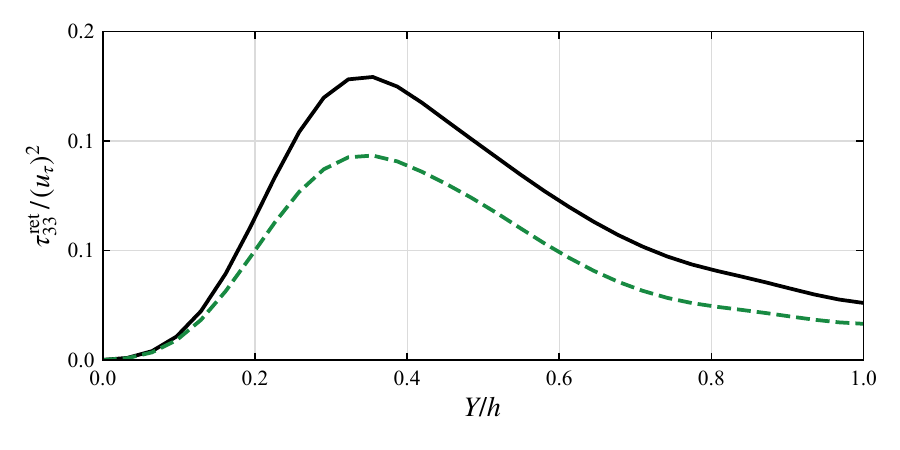}
        \caption{}
        \label{fig:pod_var_rs_w}
    \end{subfigure}

    \caption{Reynolds-stress components calculated from the reference
    (black solid line) and VAR-predicted (green dashed line)
    POD-retained fields: (a) $\tau^{\text{ret}}_{11}$;
    (b) $\tau^{\text{ret}}_{22}$; (c) $\tau^{\text{ret}}_{33}$.}
    \label{fig:pod_var_normal_rs_half_profiles}
\end{figure}

Afterwards, with the VAR-predicted POD coefficients, we reconstruct the POD-retained field, for which the Reynolds stress is evaluated. Figure~\ref{fig:pod_var_normal_rs_half_profiles} compares the three normal Reynolds-stress components ($\tau^{\text{ret}}_{11}$, $\tau^{\text{ret}}_{22}$, and $\tau^{\text{ret}}_{33}$) calculated from the reference and VAR-predicted POD-retained fields. It can be observed that the dominant component $\tau^{\text{ret}}_{11}$ is reproduced by the VAR model in close agreement with the reference. Specifically, the location of the peak is exactly captured, with only a $5.75\%$ relative error for the peak value. 
For the other two components, $\tau_{22}^{\mathrm{ret}}$ and $\tau_{33}^{\mathrm{ret}}$, the principal peak locations are reproduced exactly by the VAR model.
Only the secondary peak of $\tau_{22}^{\mathrm{ret}}$ exhibits a minor shift of $0.03h$, although some discrepancies are observed in the peak amplitudes.

\subsection{Latent Residual Results}
\label{subsec:u_residual_closure_results}

The FK-$\beta$-VAE is first applied to the POD-truncated fields, yielding a temporal sequence of latent means and variance. The input is a single-channel $64\times64$ field, and the latent dimension is set to be $d_z=24$. The FNO blocks employ a channel width of $32$ and retain $16$ Fourier modes in each spatial direction. The multi-step Koopman regularization is evaluated over a prediction sequence of four time steps. The relative weights of the Koopman and KL terms are set to $\gamma=1.0$ and $\beta=10^{-4}$, respectively. 

\begin{figure}
  \centerline{\includegraphics[width=1.2\textwidth]
  {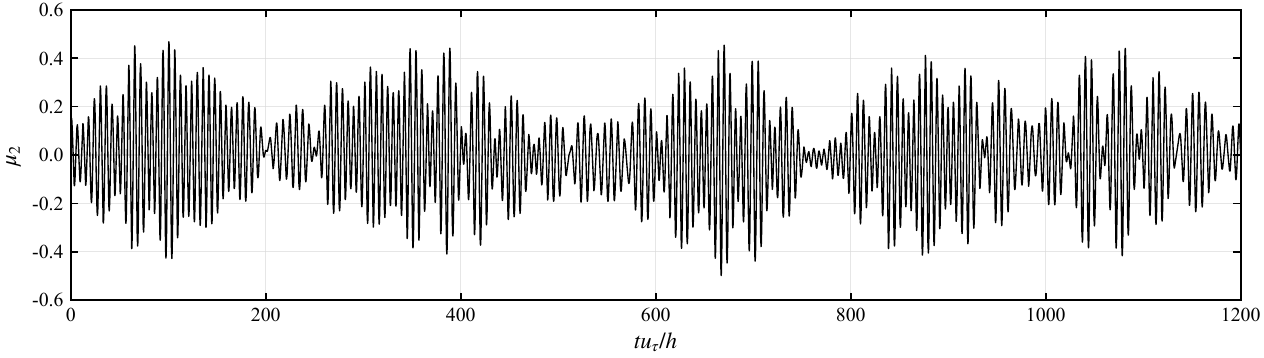}}
  \caption{Temporal sequence of the second latent mean, $\mu_2$, generated by the FK-$\beta$-VAE method.}
  \label{fig:u_eae_latent_dynamics}
\end{figure}

As one example, Figure~\ref{fig:u_eae_latent_dynamics} shows the complete temporal history of the second latent mean $\mu_2$. It can be observed that the latent variable exhibits strong temporal correlations, rather than random and uncorrelated white noise, which makes the sequential prediction feasible. 

To build the switching-VAR model describing the evolution of the latent variable, the two lag orders are set as $q^{(1)}=4$ and $q^{(2)}=32$, with an identical regularization parameter $\eta=10^{-3}$. In addition, for the current prediction horizon $18\,000\Delta t$
the switching parameters are set to $c_s=12\,000\Delta t$ and $\Delta_s=7\,000\Delta t$.

\begin{figure}
    \centering
    \captionsetup{style=capcenter}

    \begin{subfigure}{0.8\textwidth}
        \centering
        \includegraphics[width=\linewidth]
        {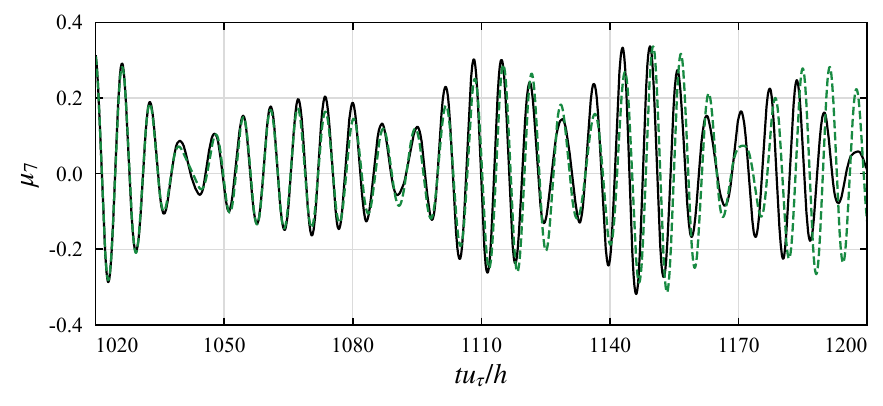}
        \caption{}
        \label{fig:u_latent_mu_dim07}
    \end{subfigure}

    \vspace{0.15em}

    \begin{subfigure}{0.8\textwidth}
        \centering
        \includegraphics[width=\linewidth]
        {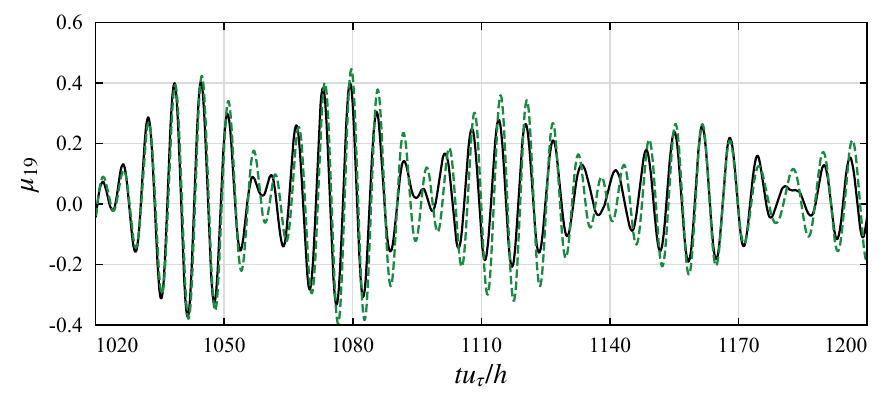}
        \caption{}
        \label{fig:u_latent_mu_dim19}
    \end{subfigure}

    \vspace{0.15em}

    \begin{subfigure}{0.8\textwidth}
        \centering
        \includegraphics[width=\linewidth]
        {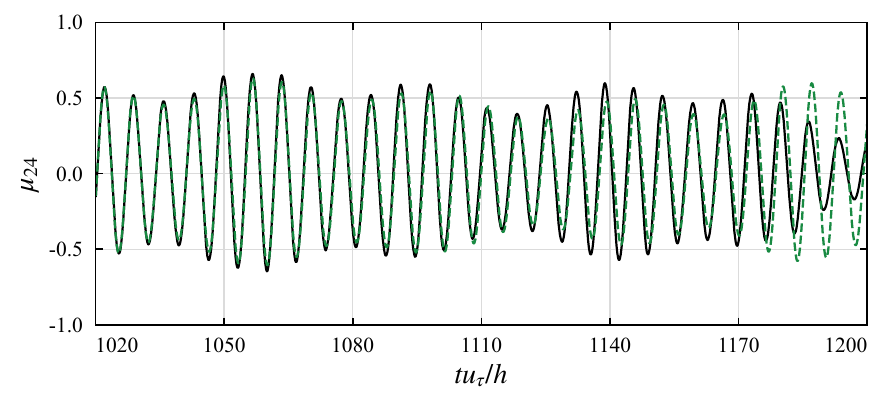}
        \caption{}
        \label{fig:u_latent_mu_dim24}
    \end{subfigure}

    \caption{Temporal sequences of latent means from the reference
    dataset (black solid line) and switching-VAR predictions
    (green dashed line): (a) $\mu_7$; (b) $\mu_{19}$; (c) $\mu_{24}$.}
    \label{fig:u_latent_rollout_dims}
\end{figure}

Figure~\ref{fig:u_latent_rollout_dims} compares the reference and predicted trajectories of three representative latent variables. It can be found that the switching-VAR model accurately predicts the overall pattern of the latent variables, closely tracking both peaks and troughs. Although predictions are more precise in the early stages and gradually diverge from the reference data over time, this error remains bounded rather than growing drastically.

\begin{figure}
    \centering
    \captionsetup{style=capcenter}

    \begin{subfigure}{0.75\textwidth}
        \centering
        \includegraphics[width=\linewidth]
        {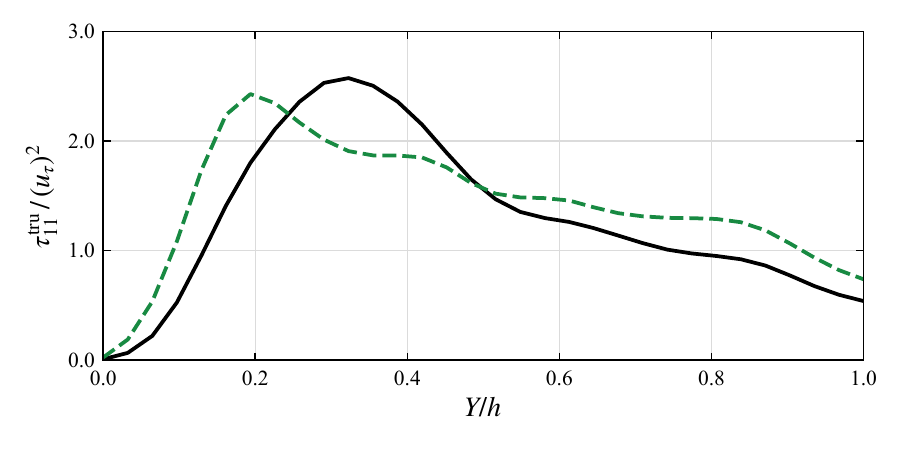}
        \caption{}
        \label{fig:residual_rs_u}
    \end{subfigure}

    \vspace{0.1em}

    \begin{subfigure}{0.75\textwidth}
        \centering
        \includegraphics[width=\linewidth]
        {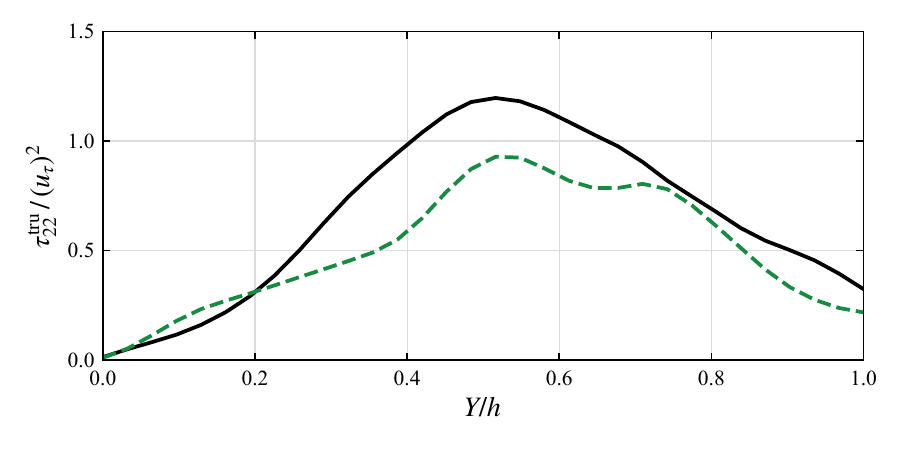}
        \caption{}
        \label{fig:residual_rs_v}
    \end{subfigure}

    \vspace{0.1em}

    \begin{subfigure}{0.75\textwidth}
        \centering
        \includegraphics[width=\linewidth]
        {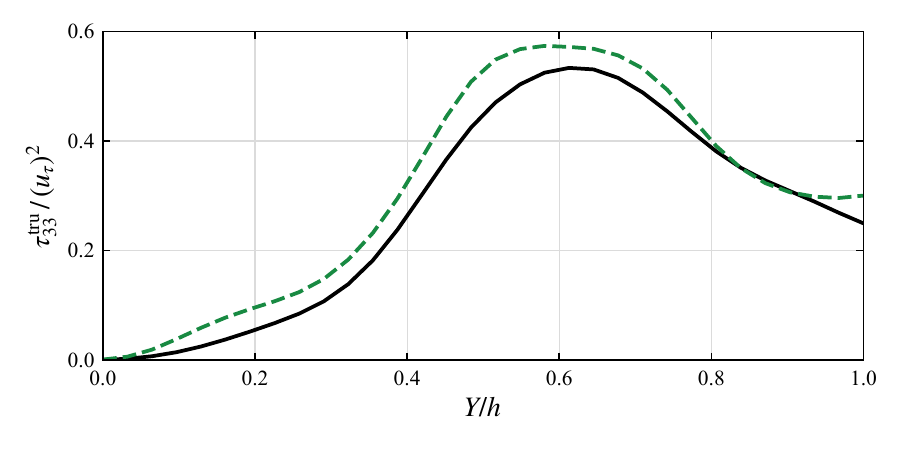}
        \caption{}
        \label{fig:residual_rs_w}
    \end{subfigure}

    \caption{Reynolds-stress components calculated from the reference
    (black solid line) and predicted (green dashed line) POD-truncated fields:
    (a) $\tau^{\text{tru}}_{11}$;
    (b) $\tau^{\text{tru}}_{22}$;
    (c) $\tau^{\text{tru}}_{33}$.}
    \label{fig:residual_normal_rs_half_profiles}
\end{figure}

Finally, the predicted latent mean, $\widehat{\boldsymbol{\mu}}_n$, is passed through the decoder to reconstruct the POD-truncated field, $\widehat{\boldsymbol{\xi}}_n^{\mathrm{tru}}$. Figure~\ref{fig:residual_normal_rs_half_profiles} compares the results of $\tau_{11}^{\mathrm{tru}}$,
$\tau_{22}^{\mathrm{tru}}$ and $\tau_{33}^{\mathrm{tru}}$, evaluated from
the reference and predicted POD-truncated fields. For $\tau_{11}^{\mathrm{tru}}$, the prediction reproduces the near-wall growth and the subsequent decay towards the outer region. Its peak amplitude is underestimated by only $5.68\%$, although the predicted peak has a shift of $0.12h$ from the reference. For $\tau_{22}^{\mathrm{tru}}$, the peak location is reproduced exactly, while its amplitude is underestimated by $22.44\%$. For $\tau_{33}^{\mathrm{tru}}$, the peak is displaced by only $0.03h$, and its amplitude is overestimated by $7.53\%$. Overall, the latent-mean switching-VAR model retains the principal spatial characteristics of Reynolds stresses from the POD-truncated field.

\subsection{Total field results}
\label{subsec:total_results}

\begin{figure}
    \centering
    \captionsetup{style=capcenter}

    \begin{subfigure}{0.75\textwidth}
        \centering
        \includegraphics[width=\linewidth]
        {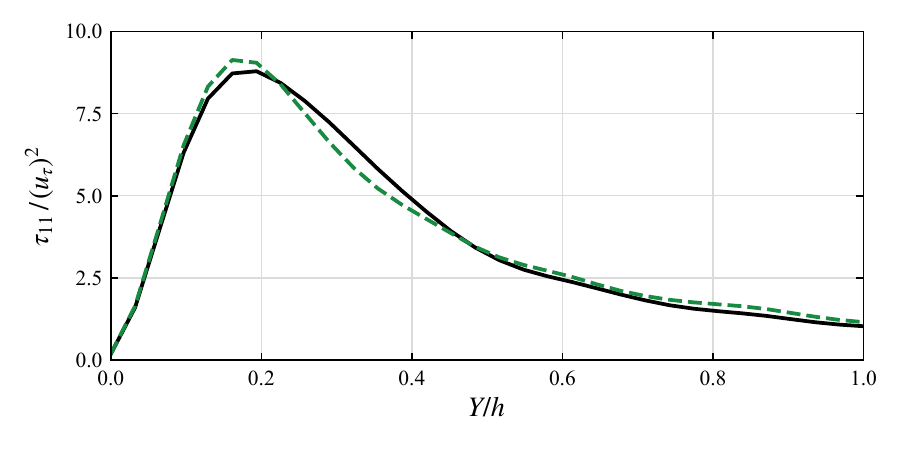}
        \caption{}
        \label{fig:total_rs_u}
    \end{subfigure}

    \vspace{0.1em}

    \begin{subfigure}{0.75\textwidth}
        \centering
        \includegraphics[width=\linewidth]
        {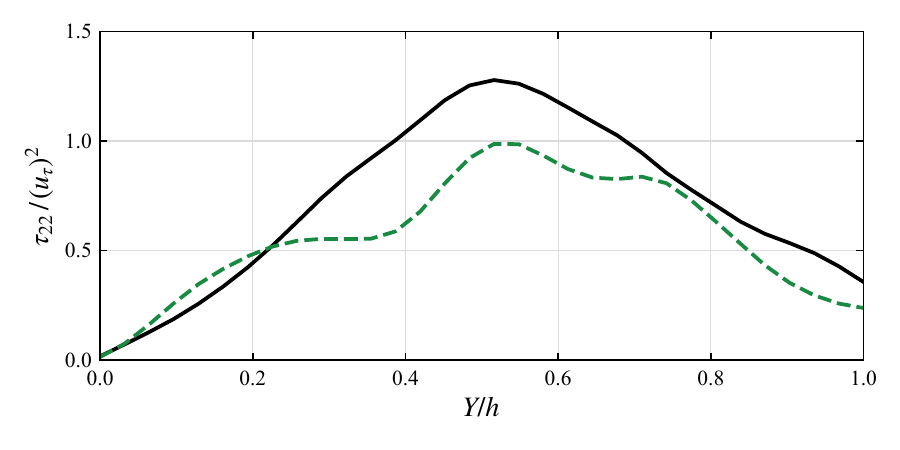}
        \caption{}
        \label{fig:total_rs_v}
    \end{subfigure}

    \vspace{0.1em}

    \begin{subfigure}{0.75\textwidth}
        \centering
        \includegraphics[width=\linewidth]
        {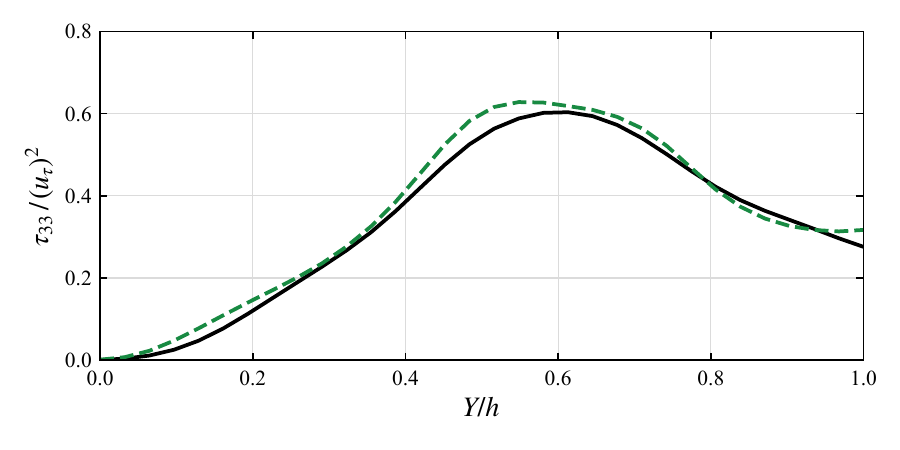}
        \caption{}
        \label{fig:total_rs_w}
    \end{subfigure}

    \caption{Reynolds-stress components calculated from the reference
    (black solid line) and hybrid-ROM-predicted (green dashed line) fields:
    (a) $\tau_{11}$;
    (b) $\tau_{22}$;
    (c) $\tau_{33}$.}
    \label{fig:total_normal_rs_profiles}
\end{figure}

\begin{figure}
  \centerline{\includegraphics[width=0.80\textwidth]
  {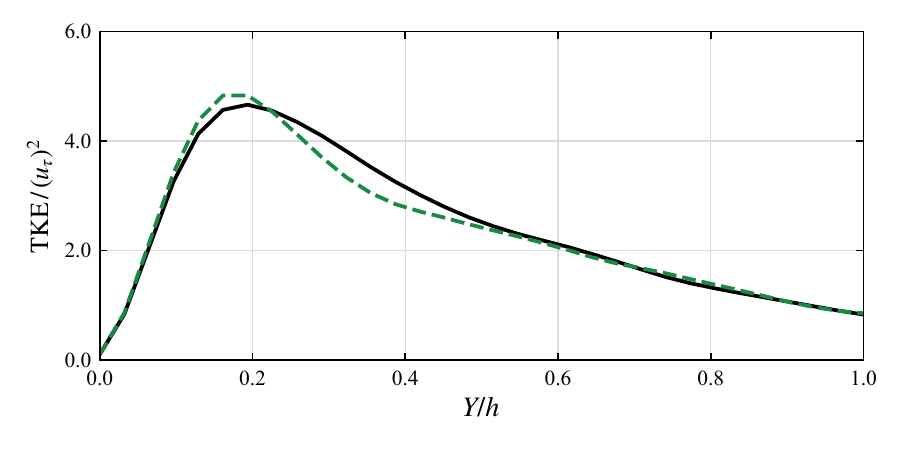}}
  \caption{Wall-normal profile of TKE from the reference
  (black solid line) and hybrid-ROM-predicted
  (green dashed line) fields.}
  \label{fig:total_tke_profile}
\end{figure}

Sequentially the full field statistics are obtained by combining the contributions from the POD-retained and POD-truncated fields,
\begin{equation}
    {\tau}_{**}
    =
    {\tau}_{**}^{\text{ret}}
    +
    {\tau}_{**}^{\text{tru}},
\end{equation}
where $*=1,~2~\text{or}~3$. Figure~\ref{fig:total_normal_rs_profiles} presents the predicted Reynolds stress components from the hybrid ROM method, in comparison with their reference, for the full field. Overall, the predicted profiles reproduce the principal spatial structures of all three components. Notably, both $\tau_{11}$ and $\tau_{33}$ are precisely reproduced, with both the general trend and amplitude exhibiting almost negligible discrepancies. For $\tau_{22}$, the deviations are slightly greater, but the proposed ROM still captures the primary spatial features to a great extent. Finally, figure~\ref{fig:total_tke_profile} compares the reference and predicted TKE profiles. The reconstructed profile closely follows the reference distribution, reproducing its near-wall growth, peak and subsequent decay towards the channel interior. The peak location is captured exactly at $Y=0.19h$, with an overestimation of only $3.61\%$ in the peak amplitude. Since the TKE combines the contributions of the three normal Reynolds-stress components, this agreement demonstrates that the combined ROM accurately recovers the principal spatial distribution of the fluctuation energy over the whole domain.

\begin{figure}
  \centerline{\includegraphics[width=0.70\textwidth]
  {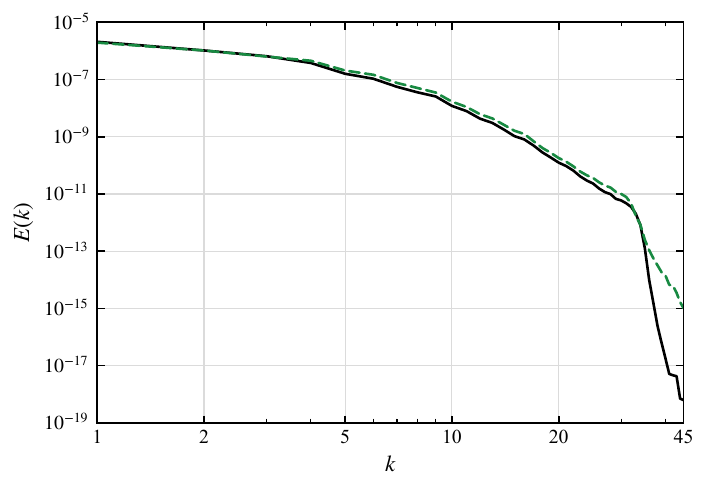}}
  \caption{Wavenumber spectrum of the total fluctuation energy from the
  reference (black solid line) and hybrid-ROM-predicted
  (green dashed line) fields.}
  \label{fig:wavenumber_spectra_comparison}
\end{figure}

To further validate the proposed hybrid ROM method, the spatial features of the reconstructed field are also assessed by evaluating the wavenumber spectrum. As shown in figure~\ref{fig:wavenumber_spectra_comparison}, the reconstructed spectrum closely follow the reference over the low and intermediate wavenumber ranges, accurately capturing the overall spectral distribution and particularly the energy decay toward smaller spatial scales. Notable differences only arise at the high wavenumber region, where the energy content is several orders of magnitude lower than other regions and therefore can be considered negligible. 

\begin{figure}
    \centering
    \captionsetup{style=capcenter}

    \begin{subfigure}{0.8\textwidth}
        \centering
        \includegraphics[width=\linewidth]
        {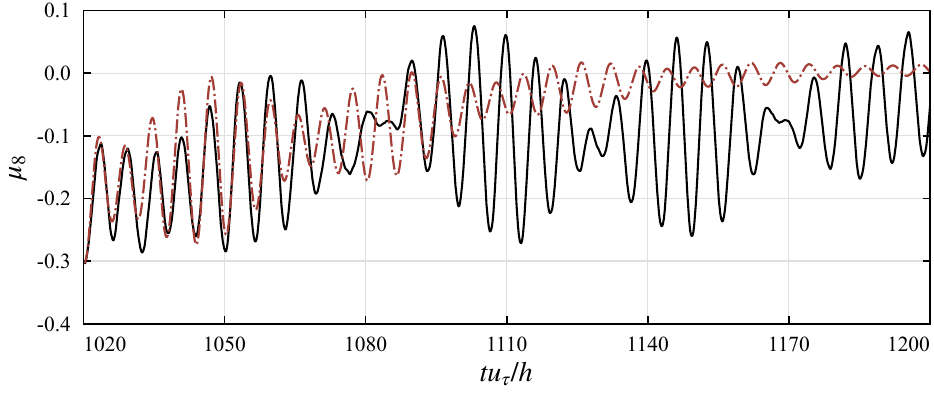}
        \caption{}
        \label{fig:u_fullfield_latent_mu_dim08}
    \end{subfigure}

    \vspace{0.8em}

    \begin{subfigure}{0.8\textwidth}
        \centering
        \includegraphics[width=\linewidth]
        {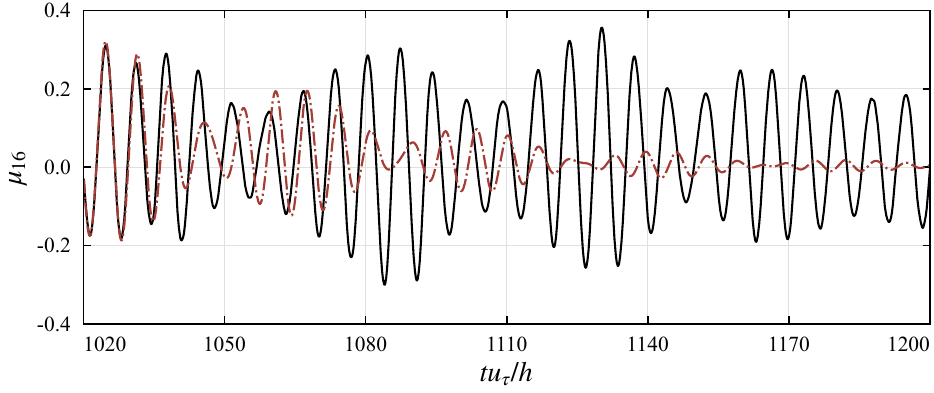}
        \caption{}
        \label{fig:u_fullfield_latent_mu_dim16}
    \end{subfigure}

    \caption{Temporal sequences of latent means from the reference dataset (black solid line) and switching-VAR predictions for the full field (dark-red dot-dash line): (a) $\mu_{8}$; (b) $\mu_{16}$.}
    \label{fig:u_fullfield_FKbetaVAE_switching_VAR_latent_dynamics}
\end{figure}

\begin{figure}
  \centerline{\includegraphics[width=0.750\textwidth]
  {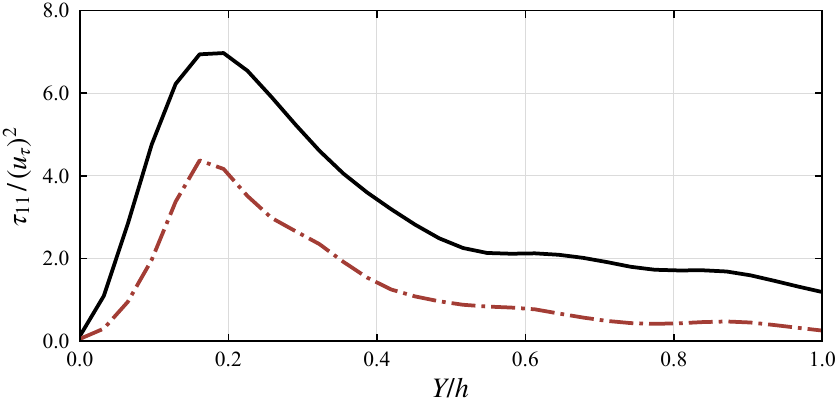}}

  \caption{Wall-normal profile of $\tau_{11}$
  from the reference (black solid line) and direct full-field
  FK-$\beta$-VAE--switching-VAR predictions
  (dark-red dash-dotted line).}
  \label{fig:u_fullfield_FKbetaVAE_switching_VAR_RS_profile}
\end{figure}

In contrast to the proposed hybrid ROM method, in which the full field is decomposed into two sub-fields through frequency-informed POD analysis, a common alternative is to compress the full field directly using an autoencoder and predicting the resulting latent variables with a temporal model \citep{MilanoKoumoutsakos2002,Murata2020,Fukami2020Hierarchical,Nakamura2021,Racca2023}. To demonstrate the necessity and advantage of the proposed field decomposition step, we further conduct a numerical experiment by directly applying the FK-$\beta$-VAE and switching-VAR models to predict the full field, in which the model parameters remain identical to those mentioned above. The corresponding predicted temporal sequences for two selected latent variables are shown in figure~\ref{fig:u_fullfield_FKbetaVAE_switching_VAR_latent_dynamics}, as well as the reference. It can be found that, without the field decomposition, the predicted latent variables can closely track the reference at the early stage. However, as the time moves forward, the results gradually deviate from the reference, exhibiting both phase mismatch and amplitude attenuation. This degradation ultimately leads to the significant misestimation of the Reynolds stresses. As one example, figure~\ref{fig:u_fullfield_FKbetaVAE_switching_VAR_RS_profile} compares the reconstructed and reference profiles for $\tau_{11}$. Although the direct full-field modeling method can still capture the approximate location of the peak and the general decay trend toward the outer region, it exhibits significant underestimation over the whole domain (up to $37.27\%$). Fundamentally, this discrepancy arises because the full field encompasses multiscale dynamics featured with distinct frequencies, which is a notable characteristic also preserved in the latent space. Therefore, capturing their long-term evolution becomes more challenging. In contrast, the proposed ROM framework decouples these multi-frequency dynamics via field decomposition, therefore enabling robust and accurate long-term predictions.

\section{Conclusions}
\label{sec:conclusions}

A hybrid POD-autoencoder ROM framework has been developed for the long-term prediction of turbulent flow statistics. Within the framework, the full field is first divided into a POD-retained field and a POD-truncated field. The retained modes are selected using a frequency-informed POD strategy based on modal energy and dominant frequency. Their POD coefficients are modeled using a VAR model. The POD-truncated field is compressed into a low-dimensional latent space using an FK-$\beta$-VAE. In this model, FNO operators help improve the spatial representation, while the Koopman operator regularizes the latent evolution process. A switching-VAR model is then used to predict the latent variables. Finally, the contributions from the two branches are combined to recover the turbulent statistics. The framework is evaluated using turbulent channel flow at $\Rey_{\tau}=110$. As indicated by the results, the VAR model captures the main statistical and spectral features of the retained POD coefficients. The latent-space model also reproduces the main dynamics and Reynolds-stress contribution of the POD-truncated field. After combining the two components, the hybrid ROM gives accurate predictions of the Reynolds stresses, TKE, and the wavenumber spectrum. In contrast, the alternative method of direct full-field modeling (i.e. without field decomposition) shows much higher prediction errors over long prediction horizons. These results demonstrate that separating flow dynamics with different temporal characteristics can improve the accuracy and robustness of long-term statistical prediction. The proposed framework thus provides a practical ROM tool for turbulent flows governed by complex, multi-scale dynamics.

\begin{bmhead}[Acknowledgements]
This research is financially supported by the Science and Technology Development Fund of Macau S.A.R. (0048/2025/ITP1, 001/2024/SKL and 0002/2025/EQP), the National Natural Science Foundation of China (52301336), and the University of Macau (MYRG-GRG2026-00121-FEG and SRG2025-00004-FST).
\end{bmhead}

\begin{bmhead}[Declaration of interests]
{The authors report no conflict of interest.}
\end{bmhead}

\bibliographystyle{jfm}
\bibliography{hybird-rom}

\end{document}